\documentclass[letterpaper,11pt]{article}
\pdfoutput=1

\usepackage{jheppub}
\usepackage{multirow}
\usepackage{shuffle}

\usepackage{breqn}

\usepackage{subfig}
\usepackage{xspace}
\usepackage[countmax]{subfloat}
\usepackage{enumitem}
\usepackage{amssymb}
\usepackage{amsmath}
\usepackage{cancel}
\usepackage{color}
\usepackage{braket}
\usepackage{graphicx}
\usepackage{multirow}
\usepackage{verbatim}
\usepackage{amsthm}
\usepackage{slashed}
\usepackage{wasysym}
\usepackage{simplewick}
\usepackage{mathtools}
\usepackage{soul}
\usepackage{xspace}

\usepackage{mathrsfs}

\def\beq{\begin{equation}}
\def\eeq{\end{equation}}
\def\bea{\begin{eqnarray}}
\def\eea{\end{eqnarray}}

\newcommand{\fd}[2]{\parbox{#1}{\includegraphics[width=#1]{#2}}}

\def\cD{\mathcal{D}}
\def\cE{\mathcal{E}}

\def\cN{\mathcal{N}}
\def\cO{\mathcal{O}}

\def\nn{{\nonumber}}

\newcommand{\Eq}[1]{Equation~\eqref{#1}}

\DeclareRobustCommand{\Sec}[1]{Sec.~\ref{#1}}

\DeclareRobustCommand{\Fig}[1]{Fig.~\ref{#1}}

\DeclareRobustCommand{\Eq}[1]{Eq.~(\ref{#1})}
\DeclareRobustCommand{\Eqs}[2]{Eqs.~(\ref{#1}) and (\ref{#2})}

\def\be{\begin{equation}}
\def\ee{\end{equation}}

\allowdisplaybreaks[3]

\newcommand{\abs}[1]{\lvert#1\rvert}

\newcommand{\ang}[1]{\langle #1 \rangle}

  \newcount\hour \newcount\minute
  \hour=\time \divide \hour by 60 \minute=\time
  \newcommand{\todaytime}{\today \ -- \number\hour :\ifnum \minute<10 0\fi\number\minute}

\def\spa#1.#2{\left\langle#1\,#2\right\rangle}
\def\spb#1.#2{\left[#1\,#2\right]}

\def\feynsl#1{
  \setbox0=\hbox{/} \setbox1=\hbox{$#1$}
  \dimen0=\wd0 \advance\dimen0 by -\wd1 \divide\dimen0 by 2
  \ifdim\wd0>\wd1 \raise.15ex\copy0\kern-\wd0\kern\dimen0\copy1\kern\dimen0
  \else \kern-\dimen0\raise.15ex\copy0\kern-\dimen0\kern-\wd1\copy1\fi}

\newskip\humongous \humongous=0pt plus 1000pt minus 100pt

\newif\ifdtup

\def\beq{\begin{equation}}
\def\eeq{\end{equation}}
\def\beeq{\begin{eqnarray}}
\def\eeeq{\end{eqnarray}}

\def\Je2e{J_{E^2 E}}

\title{Spinning Gluons from the QCD Light-Ray OPE}

\author[1]{Hao Chen,}
\author[2]{Ian Moult,}
\author[1]{and Hua Xing Zhu}

\affiliation[1]{Zhejiang Institute of Modern Physics, Department of Physics, Zhejiang University, Hangzhou, Zhejiang 310027, China}
\affiliation[2]{SLAC National Accelerator Laboratory, Stanford University, CA, 94309, USA\vspace{0.5ex}}

\emailAdd{chenhao201224@zju.edu.cn}
\emailAdd{imoult@slac.stanford.edu}
\emailAdd{zhuhx@zju.edu.cn}

\abstract{We study the transverse spin structure of the squeezed limit of the three-point energy correlator, $\langle \mathcal{E}(\vec n_1) \mathcal{E}(\vec n_2) \mathcal{E}(\vec n_3) \rangle$.  To describe its all orders perturbative behavior, we develop the light-ray operator product expansion (OPE) in QCD. At leading twist the iterated OPE of $\mathcal{E}(\vec n_i)$ operators closes onto light-ray operators $\mathbb{O}^{[J]}(\vec n)$ with spin $J$, and transverse spin $j=0,2$. We compute the $\mathcal{E}(\vec n_1) \mathcal{E}(\vec n_2)$, $\mathcal{E}(\vec n_1) \mathbb{O}^{[J]}(\vec n_2) $ and $\mathbb{O}^{[J_1]}(\vec n_1) \mathbb{O}^{[J_2]}(\vec n_2) $ OPEs as analytic functions of $J$, which allows for the description of arbitrary squeezed limits of $N$-point correlators in QCD. 
We use these results with $J=3$ to reproduce the perturbative expansion in the squeezed limit of the three-point correlator, as well as to resum the leading twist singular structure for both quark and gluon jets, including transverse spin contributions, as required for phenomenological applications.
Finally, we briefly comment on the transverse spin structure at higher twists, and show that to all orders in the twist expansion the highest transverse spin contributions are universal between quark and gluon jets, and are descendants of the leading twist transverse spin-2 operator, allowing their resummation into a simple two-dimensional Euclidean conformal block. 
Due to the general applicability of our results to arbitrary correlation functions of energy flow operators, we anticipate that they can be widely applied to improving our understanding of jet substructure at the LHC.
 }

\begin{document} 

\maketitle

\newpage

\section{Introduction}

The fundamental objects in the study of QCD at collider experiments are energy flow operators \cite{Sveshnikov:1995vi,Tkachov:1995kk,Korchemsky:1999kt,Bauer:2008dt,Hofman:2008ar,Belitsky:2013xxa,Belitsky:2013bja,Kravchuk:2018htv}
\begin{align}\label{eq:ANEC_op}
\mathcal{E}(\vec n) = \lim_{r\to \infty}  \int\limits_0^\infty dt~ r^2 n^i T_{0i}(t,r \vec n)\,.
\end{align}
Here $T_{\mu \nu}$ is the stress tensor of the theory, and $\vec n$ is a unit vector specifying the direction of the detector. The standard approach to collider experiments is the direct measurement of such operators at event level, from which one can reconstruct the energy and momentum of scattered particles, and any associated observable, on an event-by-event basis. However, from the perspective of analytically understanding energy flow from perturbative calculations, this standard approach turns out to be technically inconvenient. Event-by-event measurements of energy flow correspond to the insertion of delta function operators, $\delta(E_{\vec n} - \mathcal{E}(\vec n))$, which introduce complicated constraints on the perturbative phase space, generically preventing analytic results.  An alternative approach is to measure and compute \emph{correlation functions} of the energy flow operators over an ensemble of scattering events, $\langle \mathcal{E}(\vec n_1) \mathcal{E}(\vec n_2) \cdots \mathcal{E}(\vec n_k) \rangle$. The complete set of such correlation functions encodes the same amount of information as event-by-event observables. However, by encoding this information in correlation functions, each of which has well defined symmetry properties, information about the properties of energy flow in collisions is reorganized in a manner that makes manifest the symmetries of the underlying field theory, significantly simplifying calculations. Furthermore, in practice, correlation functions with only a few operators already provide important information on the distribution of energy flow~\cite{Chen:2020vvp}.

The one-point correlator, $\langle \mathcal{E}(\vec n)\rangle$, and two-point correlator, $\langle \mathcal{E}(\vec n_1) \mathcal{E}(\vec n_2)\rangle$, were originally proposed as event shape observables for testing QCD~\cite{Basham:1978zq,Basham:1978bw}. In recent years there has been a resurgence of interest in such observables, largely due to the observation that energy correlators admit an OPE description in terms of light-ray operators in a conformal field theory~\cite{Hofman:2008ar}. Using the light-ray operator description, a series of remarkable techniques have been developed to compute the two-point energy correlators~(EEC) and related correlators analytically to next-to-next-leading order (NNLO) in ${\cal N}=4$ super Yang-Mills~(SYM) theory~\cite{Belitsky:2013xxa,Belitsky:2013bja,Belitsky:2013ofa,Henn:2019gkr,Chicherin:2020azt}, without ever encountering infrared divergences in the intermediate steps of the calculation. At the same time, advances in techniques for multi-loop calculations have allowed the analytic calculation of the EEC at NLO in QCD~\cite{Dixon:2018qgp,Luo:2019nig,Gao:2020vyx}, and numerical calculation at NNLO~\cite{DelDuca:2016csb,DelDuca:2016ily}. There has also been substantial interest in the back-to-back limit of the EEC, which is closely related to more familiar Sudakov dynamics. Resummation of the large logarithms in this limit has been achieved at N$^3$LL~\cite{Moult:2018jzp,Ebert:2020sfi}, and generalizations to $pp$~\cite{Gao:2019ojf} and $ep$~\cite{Li:2020bub,Li:2021txc} colliders have also been proposed. The analytic results of \cite{Henn:2019gkr} have also helped to understand the structure of subleading power corrections to the EEC in the back-to-back limit~\cite{Moult:2019vou}.

Of particular recent interest at the Large Hadron Collider (LHC) is the understanding of the flow of energy at small angles within individual jets, which goes under the name of jet substructure (see \cite{Larkoski:2017jix,Asquith:2018igt,Marzani:2019hun} for reviews). Jet substructure provides new ways to study QCD at high energies, as well as novel search strategies for beyond the Standard Model physics. In terms of the correlation functions $\langle \mathcal{E}(\vec n_1) \mathcal{E}(\vec n_2) \cdots \mathcal{E}(\vec n_k) \rangle$, jet substructure is the study of the operator product expansion (OPE) limit, where the $\vec n_i$ are collinear and lie within a single jet. This perspective on jet substructure has  been proposed and developed in a recent series of papers \cite{Dixon:2019uzg,Chen:2019bpb,Chen:2020vvp,Chen:2020adz}, building on the seminal work of Hofman and Maldacena \cite{Hofman:2008ar}, as well as the works of \cite{Belitsky:2013xxa,Belitsky:2013bja,Belitsky:2013ofa}.

Formulating jet substructure as the study of the OPE limit of energy flow operators (or more generally light-ray operators \cite{Kravchuk:2018htv}), places jet substructure in a more general context, and enables the use of a wide variety of powerful techniques. In particular, light-ray operators have played a central role in many recent developments in the conformal bootstrap program, leading to their intensive study. For jet substructure, where one is interested in the OPE limit, a powerful tool is the light-ray OPE \cite{Hofman:2008ar,Kravchuk:2018htv,Kologlu:2019bco,Kologlu:2019mfz,1822249} $\mathcal{E}(\vec n_1) \mathcal{E}(\vec n_2)=\theta^{\gamma_i} \sum\mathbb{O}_i(\vec n_1)$, which allows for the expansion of products of energy flow operators over other non-local light-ray operators $\mathbb{O}_i(\vec n)$ \cite{Balitsky:1987bk,Balitsky:1988fi,Balitsky:1990ck,Braun:2003rp,Hofman:2008ar,Kravchuk:2018htv}, with the angle between the energy flow operators as the expansion parameter. This expansion is particularly convenient, since the leading terms in the expansion, which are twist-2 light-ray operators, dominate the  behavior at small angles, and are often sufficient for phenomenological purposes. The internal structure of jets is therefore determined by the spectrum of twist-2 light-ray operators in the theory, and the structure of the leading twist OPEs $\mathcal{E}(\vec n_1) \mathcal{E}(\vec n_2)$, $\mathcal{E}(\vec n_1) \mathbb{O}(\vec n_2)$ and $\mathbb{O}(\vec n_1)\mathbb{O}(\vec n_2)$, whose calculation in QCD will be a key result of this paper.

The first correlation function of energy flow operators with non-trivial shape dependence is the three-point correlator, $\langle \mathcal{E}(\vec n_1) \mathcal{E}(\vec n_2) \mathcal{E}(\vec n_3) \rangle$, which was computed in \cite{Chen:2019bpb} for both quark and gluon jets in QCD, and in $\cN=4$ super Yang-Mills (SYM). The three-point correlator provides interesting new handles to study QCD, and in particular, in the OPE or squeezed limit where two detectors are brought together,  it allows one to probe the operators appearing in the $ \mathcal{E}(\vec n_1) \mathcal{E}(\vec n_2)$ OPE, including those with non-vanishing transverse spin. In perturbation theory this transverse spin structure arises from the quantum interference of intermediate gluons in the jet, and imprints a leading twist $\cos(2\phi)$ interference pattern on detectors at infinity \cite{Chen:2020adz}.

In this paper we study in detail the structure of the OPE limit of the three-point correlator in both QCD and $\cN=4$ SYM, focusing in particular on its transverse spin structure. There are at least three motivations for this. First, the squeezed limit of the three-point correlator provides a non-trivial testing ground to develop the light-ray OPE. In this paper, we will focus primarily on the structure of the OPE at leading power in QCD, and we will compute and test  the $\mathcal{E}(\vec n_1) \mathcal{E}(\vec n_2)$, $\mathcal{E}(\vec n_1) \mathbb{O}(\vec n_2)$, and $\mathbb{O}(\vec n_1) \mathbb{O}(\vec n_2)$ OPEs, which can then be used in other more general correlators. Second,  large logarithms appear in the perturbative calculation of the squeezed limits of generic correlators, which must be resummed to derive phenomenological predictions at the LHC. Understanding the structure of these logarithms in the concrete setting of the three-point correlator allows us to develop techniques which are applicable for arbitrary correlators. Finally, the OPE limit provides significant insight into the structure of the full result, ideally enabling it to be bootstrapped.

With these motivations in mind, an outline of this paper is as follows. In \Sec{sec:param} we briefly review the three-point correlation function of energy flow operators, as well as our choice of parametrization that will be used throughout the paper. In \Sec{sec:power_expansion} we expand our results for the three-point correlator in the squeezed limit to study the structure of the twist expansion, with a particular focus on its transverse spin structure. In \Sec{sec:amp} we calculate the leading power behavior in the squeezed limit using a traditional splitting function approach to illustrate the origin of transverse spin effects in a weakly coupled gauge theory.  In \Sec{sec:ope} we present our calculations of the $\mathcal{E}(\vec n_1) \mathcal{E}(\vec n_2)$, $\mathcal{E}(\vec n_1) \mathbb{O}^{[J]}(\vec n_2) $ and $ \mathbb{O}(\vec n_1) \mathbb{O}(\vec n_2)$ OPEs in QCD and derive the leading order structure constants as analytic functions of $J$. In \Sec{sec:resum} we use the particular case of $J=3$ to perform the resummation of the squeezed limit of the three-point correlator at leading twist. We conclude in \Sec{sec:conc}.

\section{The Three-Point Energy Correlator}\label{sec:param}

Before beginning our study of the squeezed limit of the three-point correlator, we briefly review its definition following \cite{Chen:2019bpb}. This also allows us to establish our notation, and define  the parametrization of the correlator that will be used throughout the paper.

For generic angles, the three-point correlator $\ang{\mathcal{E}(\vec{n}_1)\mathcal{E}(\vec{n}_2)\mathcal{E}(\vec{n}_3)}$ is described by three respective angles. In the collinear limit, to which we restrict ourselves in this paper, this reduces to two rescaled angles and a scaling variable characterizing the overall size of the three points. For the case of a weakly coupled gauge theory, such as QCD or $\cN=4$ SYM, collinear factorization allows the leading order three-point correlator to be computed from the  $1\to 3\; (i\to a,b,c)$ splitting functions, $P_{a b c}^{(i)}$, using
\beq
\frac{1}{\sigma_{\rm tot}}\frac{d^3\Sigma_i}{d x_1 d x_2 d x_3}=\sum_{a,b,c} \int d\Phi_c^{(3)} \frac{4 g^4}{s_{123}} P_{a b c}^{(i)} \frac{E_a E_b E_c}{(Q/2)^3} \delta\left(x_1-\frac{s_{a b}}{4 E_a E_b}\right) \delta\left(x_2-\frac{s_{b c}}{4 E_b E_c}\right)\delta\left(x_3-\frac{s_{a c}}{4 E_a E_c}\right)\,.
\eeq
Here $\Phi_c^{(3)} $ is the three-particle collinear phase space, and the detector positions are enforced by the delta functions, with
\begin{align}
x_1 = \frac{1 - \cos \theta_{23}}{2} \simeq \frac{\theta_{23}^2}{4}\,,\qquad  x_2\simeq \frac{\theta_{13}^2}{4}\,, \qquad x_3\simeq\frac{\theta_{12}^2}{4}\,.
\end{align} 
Here we normalize to the jet energy $Q/2$ as in \cite{Chen:2020adz}, rather than the total energy $Q$ in $e^+e^-$ to dijet as in \cite{Chen:2019bpb}.

As shown in \cite{Chen:2020adz}, in the collinear limit the leading order dependence on the overall size and shape of the triangle factorizes. We will show how this arises from the light-ray OPE perspective later in this paper. To manifest this factorization, we introduce the maximal length $x_L$ as a scaling variable, and a complex cross-ratio variable $z$ to describe the shape. If we order the sides, $\{x_1,x_2,x_3\}$, of the triangle formed by the correlator as $x_S\leq x_M\leq x_L$ then $z\bar{z}=x_M/x_L,\;(1-z)(1-\bar{z})=x_S/x_L$. The result for the three-point energy correlator can then be written as
\beq\label{eq:three_point_general}
\frac{1}{\sigma_{\rm tot}}\frac{d^3\Sigma_{i}}{dx_L \, d \mathrm{Re}{z}\, d\mathrm{Im}{z}}=8\times\frac{1}{\pi} \left(\frac{\alpha_s}{4\pi}\right)^2  \frac{16}{x_L} G_{i}(z)\,,
\eeq
where the factor $8$ comes from normalizing to $Q/2$.
The function $G_i(z)$ describes the dependence on the cross-ratio variable, and depends both on the underlying theory, as well as the particle type initiating the jet, e.g. quark or gluon for the case of QCD. The function $G_i(z)$ was computed for quark and gluon jets in QCD, and in $\cN=4$ SYM in \cite{Chen:2019bpb}. For example, in $\cN=4$ SYM, the result takes the compact form\footnote{Here we have slightly simplified the result as compared to its form in \cite{Chen:2019bpb}, by replacing higher powers of $(z-\bar z)$ in denominators with derivatives acting on $\Phi(z)$, using identities inspired by \cite{Chicherin:2020azt}.}
\begin{align}\label{eq:three_point_N4}
&G_{\cN=4}(z)=\frac{1+u+v}{2uv}(1+\zeta_2)-\frac{1+v}{2uv}\log(u)-\frac{1+u}{2uv}\log(v)\nn \\
&-(1+u+v)(\partial_u +\partial_v)\Phi(z)+\frac{(1+u^2+v^2)}{2uv}\Phi(z) +\frac{(z-\bar z)^2(u+v+u^2+v^2+u^2v+uv^2)}{4u^2 v^2}\Phi(z)\nn \\
&+\frac{(u-1)(u+1)}{2u v^2}D_2^+(z) +\frac{(v-1)(v+1)}{2u^2 v}D_2^+(1-z)+\frac{(u-v)(u+v)}{2uv}D_2^+\left( \frac{z}{z-1} \right)\,,
\end{align}
where $u=z\bar z$, $v=(1-z)(1-\bar z)$,
\begin{align}
  \label{eq:1}
  \Phi(z) =\frac{2}{z-\bar z} \left(  {\rm Li}_2(z) - {\rm Li}_2(\bar z) + \frac{1}{2} \left(\log(1-z) - \log(1 - \bar z) \right) \log (z \bar z) \right)\,,
\end{align}
is the standard box function, and
\begin{align}
  \label{eq:2}
  D_2^+(z) = {\rm Li}_2(1 - |z|^2) + \frac{1}{2} \log(|1-z|^2) \log(|z|^2)\,,
\end{align}
is a weight two function even under $z \leftrightarrow \bar z$. The results for quark and gluon jets in QCD are slightly longer, but can expressed in terms of the same class of functions.

In this parametrization, the squeezed/OPE limit corresponds to $z\to 0$ or equivalently $z\to 1$. Despite the simplicity of the final result, these limits are quite rich, and provide much information on the structure of the energy correlators.

Before proceeding, we note that relating $\frac{1}{\sigma_{\rm tot}}\frac{d\Sigma_{i}}{dx_L \, d \mathrm{Re}{z}\, d\mathrm{Im}{z}}$ to the fully differential collinear triple energy correlation $\ang{\mathcal{E}(\vec{n}_1)\mathcal{E}(\vec{n}_2)\mathcal{E}(\vec{n}_3)}$ requires several changes of variables that introduce numerical factors that we record here for completeness. First, $\frac{1}{\sigma_{\rm tot}}\frac{d\Sigma_{i}}{dx_L \, d \mathrm{Re}{z}\, d\mathrm{Im}{z}}$ is defined by integrating out the overall rotation around the jet axis, which introduces a factor of $2\pi$ since $\ang{\mathcal{E}(\vec{n}_1)\mathcal{E}(\vec{n}_2)\mathcal{E}(\vec{n}_3)}$ is rotationally invariant when the collinear source is unpolarized. Second, $\mathbb{Z}_2$ symmetry $z\leftrightarrow 1-z$ contributes another factor of $2$. For concreteness, suppose $\theta_{13}>\theta_{12},\theta_{23}$, that is $\displaystyle{x_L=\frac{\theta_{13}^2}{4}}$. In the collinear limit, the integration measure $\displaystyle{ d^2\vec{n}_1d^2\vec{n}_2 d^2\vec{n}_3}$, after gauge fixing the overall translation and rotation, is
\beq
\theta_{13} d\theta_{13} d^2\vec{n}_2 =\theta_{13} d\theta_{13}\; (\theta_{13}^2 d\mathrm{Re} z\; d\mathrm{Im} z)=8 x_L dx_L d\mathrm{Re} z\; d\mathrm{Im} z.
\eeq
Therefore, we establish the relation 
\beq \label{eq: conversion_relation}
\frac{1}{x_L}\frac{1}{\sigma_{\rm tot}}\frac{d^3\Sigma_{i}}{dx_L \, d \mathrm{Re}{z}\, d\mathrm{Im}{z}}=32\pi \ang{\mathcal{E}(\vec{n}_1)\mathcal{E}(\vec{n}_2)\mathcal{E}(\vec{n}_3)}.
\eeq

\section{The Twist Expansion in the Squeezed Limit}\label{sec:power_expansion}

In this section we use the results for the integrated three-point correlator calculated in \cite{Chen:2019bpb} to investigate the structure of the power expansion in the squeezed limit, and present the expansion to twist-8 for unpolarized quark and gluon jets, and unpolarized jets in $\mathcal{N}=4$ SYM. We will then understand the leading twist behavior both from the perspective of standard splitting functions in \Sec{sec:amp} and from the light-ray OPE in \Sec{sec:ope}.

Although the primary focus of this paper is on understanding the leading twist behavior of the squeezed limit, we also make several basic comments about features of the higher twist behavior that we think would be interesting to explore further. While the light-ray OPE has been well tested for the OPE between two light-ray operators in the presence of local operators, the OPE of two light-ray operators in the presence of additional light-ray operators has not been previously tested. The perturbative data generated from the expansion of the three-point correlator therefore provides valuable data for testing the light-ray OPE in this setup. 

\begin{figure}
\begin{center}
\includegraphics[width=10cm]{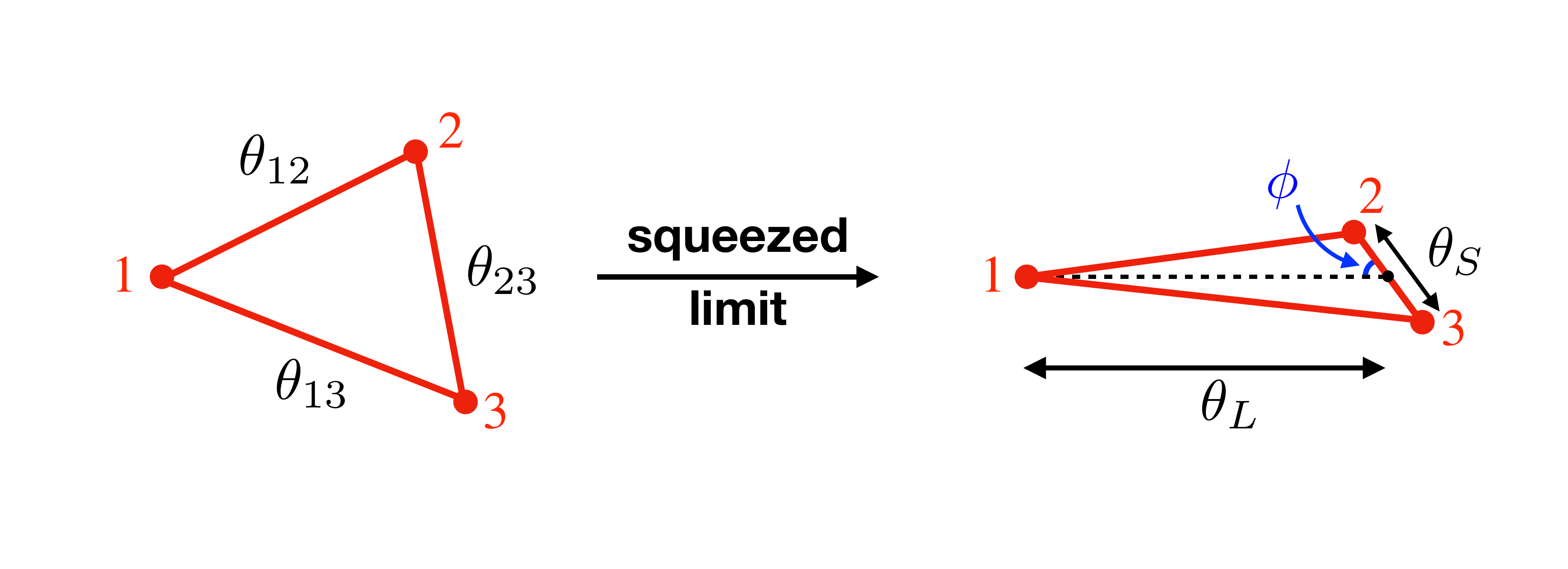}
\caption{Symmetric parametrization of the squeezed limit, as described in the text. This parametrization ensures that linear power corrections associated with an asymmetric parametrization are absent.}
\label{squeezed_fig}
\end{center}
\end{figure}

We consider the OPE limit where two detectors, which we take to be detectors $2$ and $3$, are squeezed with respect to the third detector. In other words, $\theta_{23}\ll \theta_{13}\sim \theta_{12}$, as shown in Fig.~\ref{squeezed_fig}. In the $z, \bar z$ parametrization given above, this corresponds to either the $z\to 0$ or $z\to 1$ limits. These are related by symmetry, and here for concreteness we take $z\to 1$. Instead of directly expanding with respect to the $z$ variable, we consider a more symmetric configuration where the expansion point is chosen to be the midpoint of the squeezed pair. This corresponds to the variable change $(x_L, z)\to(x, w)$ where
\beq
x_L=\frac{x}{(1+w)(1+w^*)},\quad z=\frac{1-w}{1+w}\,.
\eeq
For the overall cross section, we obtain the relation
\bea
\frac{1}{\sigma_{\rm tot}}\frac{d^3\Sigma_i}{d x\; d \mathrm{Re} w \;d\mathrm{Im}w}&=&\frac{4}{(1+w)^3 (1+w^*)^3}\times \frac{1}{\sigma_{\rm tot}}\frac{d^3\Sigma_i}{d x_L d \mathrm{Re} z \;d\mathrm{Im}z}
\nn
\\
&=&8\times \frac{1}{\pi}\left(\frac{\alpha_s}{4\pi}\right)^2 \frac{16}{x} \left[\frac{4}{(1+w)^2(1+w^*)^2}G_i(z)\right]\bigg\vert_{z\to\frac{1-w}{1+w}}\,.
\eea
The variables $(x, w)$ have simple relations with $(\theta_L, \theta_S,\phi)$ shown in Fig.~\ref{squeezed_fig} 
\beq
x=\frac{\theta_L^2}{4},\quad \abs{w}=\frac{\theta_S}{2\theta_L}, \quad \arg{w}=\phi.
\eeq
The advantage of using the variables $(x, w)$ is that the expansion in the squeezed limit ($\abs{w}\ll 1$) of $\displaystyle \widetilde{G}_i(w)= \left[\frac{4}{(1+w)^2(1+w^*)^2}G_i(z)\right]\bigg\vert_{z\to\frac{1-w}{1+w}}$ only contains even powers of $\abs{w}$. The original asymmetric parameterization in $z$ induces corrections at odd orders in the expansion.

\subsection{Integrand Level Expansion}\label{sec:expansion}

Although we can obtain the expansion in the squeezed limit by directly expanding the fully integrated correlator of \cite{Chen:2019bpb}, it is also useful to be able to extract the twist expansion directly by expanding at the level of the integrand. This allows us to make a correspondence between structures appearing in the result and momentum regions in the perturbative calculation. We believe that this is particularly interesting for comparisons with the light-ray OPE, as it may provide insight into how the light-ray OPE organizes perturbative calculations. Furthermore, it allows us to obtain OPE data without computing the full integrated result, which may also be useful in bootstrap approaches.

In terms of $(x_L,z,\bar{z})$ variables, squeezing particle 2 and 3 corresponds to taking the limit $z\to1$. Correspondingly, we parametrize $z,\bar{z}$ by $z=1-r \exp(i \vartheta)$ in this subsection. For generic values of the momentum fractions, we can directly expand the integrand (around $r=0$) first and then integrate over momentum fractions.  However, the main complication arises from the factors of $1/s_{123}$ or $1/s_{123}^2$ in the integrand when particle 1 becomes soft $(E_1/Q\lesssim\abs{1-z}^2)$ since
\beq
s_{123}=x_L\left( E_1\left( E_2+ E_3 \abs{z}^2\right)+E_2 E_3 \abs{1-z}^2 \right).
\eeq
In this case (when $E_1/Q,\abs{1-z}\ll 1$), we can first employ the expansions
\bea
\frac{1}{s_{123}}&=&\sum_{i=0}^{\infty} \frac{\left(E_1^2 \abs{z}^2+E_1 E_2\abs{1-z}^2\right)^i}{(E_1(E_2+\abs{z}^2(Q/2-E_2))+\abs{1-z}^2 E_2(Q/2-E_2))^{i+1}}\,,\\
\frac{1}{s_{123}^2}&=&\sum_{i=0}^{\infty} \frac{(i+1)\left(E_1^2 \abs{z}^2+E_1 E_2\abs{1-z}^2\right)^i}{(E_1(E_2+\abs{z}^2(Q/2-E_2))+\abs{1-z}^2 E_2(Q/2-E_2))^{i+2}}\,, \label{eq: soft_end_exp}
\eea
where we have used $E_3=Q/2-E_1-E_2$ and later will set $Q/2=1$ for convenience. 

We illustrate the procedure with the integral
\beq
I=\int_{0}^{1}dE_2 \int_{0}^{1-E_2}  dE_1 \frac{E_1^2 E_2^2 E_3^2}{s_{123}^2(E_1+E_2)^2}\Bigg|_{E_3\to1-E_1-E_2}\,.
\eeq
To treat the soft limit of the particle 1, it is standard to separate the integration region of $E_1$ to $\mathcal{R}_1=(0,\delta)$ and $\mathcal{R}_2=(\delta,1-E_2)$ so that in $\mathcal{R}_2$ we can first expand the integrand around $r=0$ before integration, while in $\mathcal{R}_1$ we can make use of expansion (\ref{eq: soft_end_exp}) and then expand around $r=0$ after the $E_1$ integration. This leaves us with the two integrals
\begin{align}
I_1(\delta)&=\!\! \int_{0}^{1}\!\! dE_2 \!\! \int_{0}^{\delta}\!\! dE_1 \frac{E_1^2 E_2^2 (1-E_1-E_2)^2}{s_{123}^2(E_1+E_2)^2} \\
&=\!\! \int_{0}^{1}\!\! dE_2 \!\! \int_{0}^{\delta}\!\! dE_1\frac{E_1^2 E_2^2 (1-E_1-E_2)^2 (E_1+E_2)^{-2}}{ (E_1(E_2+\abs{z}^2(1-E_2))+\abs{1-z}^2 E_2(1-E_2))^{2}}+\cdots \nonumber\\
I_2(\delta)&=\!\! \int_{0}^{1}\!\! dE_2 \!\! \int_{\delta}^{1-E_2}\!\!\!\!\!\!\!\!\!\!\!\! dE_1 \frac{E_1^2 E_2^2 (1-E_1-E_2)^2}{s_{123}^2(E_1+E_2)^2}\,,\nn \\
&= \!\! \int_{0}^{1}\!\! dE_2 \!\! \int_{\delta}^{1-E_2}\!\!\!\!\!\!\!\!\!\!\!\! dE_1 \frac{E_2^2(1-E_1-E_2)^2}{(E_1+E_2)^2(1-E_1)^2}\left[1+ 4 r \cos\vartheta\frac{1-E_1-E_2}{1-E_1}+\cdots\right] \nonumber \,.
\end{align}
Since the sum of $I_1(\delta)$ and $I_2(\delta)$ is independent of $\delta$, we can consider $\delta \to 0$ limit to further simplify the expansion result, 
\bea
\lim_{\delta \to 0}I_1(\delta) \!&=&\! \frac{1}{300} r^2(-30 \log \delta +60 \log r-37) +\frac{1}{150} r^3 \cos \vartheta (-60 \log \delta+120 \log r-59)+\cdots\,,\nonumber\\
\lim_{\delta \to 0}I_2(\delta) \!&=&\!  \pi ^2-\frac{59}{6}+\frac{r}{6} \cos\vartheta(119-12 \pi ^2) +\frac{r^2}{30} (3 \log \delta+3 \cos 2 \vartheta+10)\nn \\
&+&\frac{r^3}{30}  \cos \vartheta (12 \log \delta+6 \cos 2 \vartheta+35)+\cdots\,.
\eea
As expected, the $\delta$ dependence cancel in the sum. 
An interesting feature of this expansion, that we will return to in the following section, is the appearance of $\log r$ ($\log 2\abs{w}$) and $\pi^2$ at higher twists in the squeezed limit. By performing the expansion by momentum regions, we can see that the $\log r$ arise from the soft limit of the unsqueezed particle, whereas the $\pi^2$ arises from the soft limit of the squeezed pair.

\subsection{Quark and Gluon Jets in QCD}\label{sec:QCD_power_discussion}

In this subsection we present the expanded results for QCD jets.
We begin by considering the case of quark and gluon jets in QCD. Although they are considerably more complicated than the case of $\cN=4$ SYM, this actually turns out to be advantageous, as they exhibit a more general transverse spin structure, which is absent in the $\cN=4$  SYM result due to supersymmetric cancellations.

We find to $\cO(|w|^4)$ for quark jets
\begin{align}\label{eq:quark_result_expand}
&\widetilde{G}_q(w)= C_F n_f T_f\left\{
\frac{1}{\left| w\right| ^2}\left[ \frac{13}{4800}-{\color{blue}\frac{1}{720} \cos (2 \phi )} \right]
+\left[{\color{blue}-\frac{\cos (4\phi )}{1680}}+\frac{37 \cos (2 \phi )}{10080}-\frac{\pi ^2}{3}+\frac{111199}{33600}\right]\right. \nn\\
& \left. +\left| w\right| ^2 \left[ {\color{magenta}-\frac{67}{105} \log(2 \left| w\right| )} {\color{blue}-\frac{\cos (6 \phi )}{3024}} +\frac{331 \cos (4 \phi )}{151200} -\left(\frac{46}{3} \pi ^2-\frac{22853623}{151200}\right) \cos (2 \phi )-\frac{28 \pi^2}{3}+\frac{21468341}{235200}\right] \right.\nn\\
& +\left| w\right|^4 \left[ {\color{magenta}\left(-\frac{2996}{495} \cos (2 \phi )-\frac{12317}{1155}\right) \log (2\left| w\right| )} {\color{blue}-\frac{\cos (8 \phi )}{4752}}+\frac{53\cos (6 \phi )}{36960} -\left(110 \pi ^2 -\frac{32817971}{30240}\right)\cos (4 \phi ) \right.\nn\\
&\left. \left.\qquad\quad-\left(\frac{464}{3} \pi ^2- \frac{83514316033}{54885600}\right) \cos (2 \phi )-83 \pi^2+\frac{41787326893}{51226560}\right]
 \right\}\nn \\
 &+C_F^2\left\{
 \frac{1}{20 \left| w\right| ^2}+\left[\frac{13}{180} \cos (2\phi )+\frac{59}{360}\right]\right.\nn\\
&+\left| w\right| ^2 \left[{\color{magenta}-\frac{4}{21} \cos (2 \phi) \log (2 \left| w\right| )}+\frac{23}{350} \cos (4\phi )+\frac{6529 \cos (2 \phi )}{22050}-\frac{29}{3150}\right]\nn\\
& + \left| w\right| ^4 \left[{\color{magenta}\left(-\frac{208}{315} \cos (2 \phi )-\frac{88}{315} \cos (4\phi )-\frac{4}{15}\right) \log (2 \left| w\right| )}\right.\nn\\
&\left.  \left.  +\frac{44153 \cos (2 \phi)}{396900}+\frac{102983 \cos (4 \phi )}{396900}+\frac{79 \cos (6 \phi)}{1260}+\frac{6613}{37800}\right]
 \right\}\nn\\
 &+C_A C_F\left\{
 \frac{1}{\left| w\right| ^2}\left[ {\color{blue} \frac{\cos (2 \phi )}{1440}} +\frac{91}{4800}\right]
  +\left[\frac{1}{6} \pi ^2 \cos (2 \phi )-\frac{26779
   \cos (2 \phi )}{16800}+{\color{blue}\frac{\cos (4 \phi )}{3360}}+\frac{3 \pi
   ^2}{8}-\frac{41443}{11200}\right]\right.\nn\\
   &  +\left| w\right| ^2
   \left[{\color{magenta}\left(\frac{13}{210} \cos (2 \phi )-\frac{26}{105}\right) \log (2 \left| w\right| )}
   +{\color{blue}\frac{\cos (6 \phi )}{6048}}
   +\left(\frac{5}{3} \pi ^2 -\frac{354143}{21600}\right) \cos (4 \phi )\right.\nn\\
&\qquad\qquad\left.  +\left(\frac{115}{12} \pi ^2 -\frac{99934619}{1058400}\right) \cos (2 \phi )
   +\frac{41 \pi^2}{6}-\frac{47421299}{705600}\right]\nn\\
& +\left| w\right| ^4
\left[{\color{magenta}\left(\frac{8968 \cos (2 \phi )}{3465}+\frac{4}{45} \cos (4 \phi)+\frac{6553}{2310}\right) \log (2 \left|w\right| )}
+{\color{blue}\frac{\cos (8 \phi )}{9504} }
+\left(\frac{59}{6} \pi ^2 -\frac{23046257}{237600}\right) \cos (6 \phi)\right.\nn\\
&\quad \left. \left. +\left(\frac{653}{12} \pi ^2-\frac{243477673}{453600}\right) \cos (4 \phi)
+\left(\frac{185}{2} \pi ^2 -\frac{349888112069}{384199200}\right) \cos (2 \phi)
+\frac{383 \pi^2}{8}-\frac{120459334379}{256132800}\right]
 \right\}\nn\\
 &+\left(C_F^2-\frac{1}{2}C_A C_F\right)\left\{
 \left[ \left(-\frac{13}{6} \pi ^2 +\frac{38483}{1800}\right) \cos (2 \phi )-\frac{17 \pi ^2}{12}+\frac{50339}{3600}\right]\right.\nn\\
& +\left| w\right| ^2 \left[{\color{magenta}-\frac{4}{105} \log(2 \left| w\right| )}
-\left(15 \pi ^2-\frac{1865287}{12600}\right) \cos (4 \phi )
-\left(\frac{64}{3} \pi ^2- +\frac{66334}{315}\right) \cos (2 \phi )
-\frac{73 \pi^2}{6}+\frac{588211}{4900}\right]\nn\\
&  + \left| w\right| ^4 \left[ {\color{magenta}\left(\frac{68}{315} \cos (2 \phi )-\frac{4}{15}\right) \log (2\left| w\right| )}
-\left(\frac{343}{6} \pi ^2-\frac{10663507}{18900}\right) \cos (6 \phi )\right.\nn\\
&\qquad \left.  \left.-\left(\frac{407}{6} \pi ^2-\frac{117168}{175} \right) \cos (4 \phi )
-\left(\frac{503}{6} \pi ^2- +\frac{82101599}{99225}\right) \cos (2 \phi )
-\frac{173\pi ^2}{4}+\frac{32276071}{75600}\right]
 \right\}\nn\\
 &+\cO(|w|^6)\,.
\end{align}
Similarly, for gluon jets we find
\beq\label{eq:gluon_result_expand}
\begin{split}
&\widetilde{G}_{g}(w)=C_F n_f T_F\left\{
 \frac{3}{320\left| w\right| ^2}+\left[\frac{19}{800} \cos (2 \phi )-\frac{13}{1600}\right]\right.\\
&+\left| w\right|^2
 \left[{\color{magenta}-\frac{3}{7} \log (2 \left| w\right| )}
 +\frac{137 \cos (4 \phi )}{5600}
 -\frac{39 \cos (2 \phi)}{5600}
 -\frac{11937}{78400}\right]\\
&\left. +\left| w\right| ^4 
\left[{\color{magenta}\left(-\frac{4}{5} \cos (2 \phi )-\frac{67}{35}\right) \log (2\left| w\right| )}
+\frac{83 \cos (6 \phi )}{3360}-\frac{13 \cos (4 \phi)}{3360}
-\frac{2453 \cos (2 \phi )}{7200}+\frac{245579}{705600}\right]
\right\}\\
&+C_A n_f T_F\left\{
\frac{1}{\left| w\right| ^2}\left[\frac{7}{800}{\color{blue}-\frac{1}{720} \cos (2 \phi )} \right]
+\left[{\color{blue}-\frac{\cos (4 \phi )}{1680} }+\left(\frac{5}{6} \pi ^2-\frac{207251}{25200}\right) \cos (2 \phi )+\frac{\pi^2}{4}-\frac{120899}{50400}\right]\right.\\
& +\left| w\right| ^2 \left[{\color{magenta}\left(-\frac{8}{315}\cos (2 \phi )-\frac{22}{63}\right) \log (2 \left| w\right| )}
{\color{blue}-\frac{\cos (6 \phi )}{3024}}\right. \\
&\qquad\quad \left.+\left(\frac{53}{3} \pi ^2 -\frac{13182121}{75600}\right) \cos (4 \phi)
+\left(\frac{43}{3} \pi ^2 -\frac{224661347}{1587600}\right) \cos(2 \phi )
+\frac{25 \pi^2}{6}-\frac{5317297}{127008}\right]\\
&+\left| w\right| ^4
   \left[{\color{magenta}\left(-\frac{17588 \cos (2 \phi )}{3465}-\frac{122 \cos (4 \phi)}{3465}-\frac{964}{105}\right) \log (2 \left| w\right| )}
   {\color{blue}-\frac{\cos (8 \phi )}{4752}}
   +\left(\frac{799}{6} \pi ^2 -\frac{1092978311}{831600} \right)\cos (6 \phi)\right. \\
&\left.\left.   +\left(\frac{961}{6} \pi ^2-\frac{303716037469}{192099600} \right)\cos (4 \phi)
   +\left(\frac{511}{6} \pi ^2-\frac{162328436041}{192099600}\right) \cos (2\phi )
   +\frac{123 \pi^2}{4}-\frac{714502853}{2328480}\right]
\right\}\\
&+C_A^2\left\{
\frac{1}{\left| w\right| ^2}\left[{\color{blue}\frac{\cos (2 \phi )}{1440}}+\frac{49}{800}\right]
+\left[{\color{blue}\frac{\cos (4 \phi )}{3360}} -\left(\frac{4}{3} \pi ^2-\frac{66881}{5040}\right) \cos (2 \phi )-\frac{5 \pi ^2}{8}+\frac{318193}{50400}\right]\right.\\
&+\left| w\right| ^2
   \left[{\color{magenta}\left(-\frac{73}{630} \cos (2 \phi )-\frac{62}{315}\right) \log (2 \left| w\right| )}
   +{\color{blue}\frac{\cos (6 \phi )}{6048}}\right.\\
&\qquad\quad \left.   -\left(\frac{44}{3} \pi ^2 - \frac{10951547}{75600}\right)\cos (4 \phi )
   -\left(\frac{191}{12} \pi ^2-\frac{125023781}{793800}\right) \cos (2 \phi )
   -6 \pi^2+\frac{7550981}{127008}\right]\\
& +\left| w\right| ^4
   \left[{\color{magenta}\left(\frac{964}{495} \cos (2 \phi )-\frac{599 \cos (4 \phi)}{3465}+\frac{557}{210}\right) \log (2 \left| w\right| )}
   +{\color{blue}\frac{\cos (8 \phi )}{9504}}
   -\left(\frac{256}{3} \pi ^2 -\frac{28018675 }{33264}\right) \cos (6 \phi)\right. \\
&\quad\left.\left. -\left(\frac{1375}{12} \pi ^2-\frac{217341879809 }{192099600} \right)\cos (4 \phi)   
   -\left(\frac{208}{3} \pi ^2-\frac{18842665223}{27442800}\right) \cos (2\phi )
   -\frac{245 \pi^2}{8}+\frac{3544660961}{11642400}\right]
\right\}\\
&+\cO(|w|^6)\,.
\end{split}
\eeq
Higher order terms can be obtained from the full result in \cite{Chen:2019bpb}, but the expansion to this order is sufficient to illustrate the most interesting patterns.

Due to the relatively complicated structure of these results, we have highlighted two particular classes of contributions to guide the eye of the reader. First, in blue we have highlighted the terms with highest transverse spin at each power in the twist expansion, namely $\cos 2n \phi$ at order $\abs{w}^{2n-4}$. Recall  that a $\abs{w}^{\tau-4}$ dependence corresponds to a (collinear) twist-$\tau$ operator, while a $\cos 2 j \phi$ dependence corresponds to a transverse spin-$2j$ operator. Only even transverse spins contribute due to the bosonic symmetry of the energy operators of the squeezed pair. Second, in magenta, we have highlighted terms with a logarithmic dependence on the OPE parameter $|w|$, which we believe have an interesting connection to endpoint divergences in gauge theory calculations, as we will discuss shortly.

\begin{figure}
\begin{center}
\subfloat[]{
\includegraphics[scale=0.18]{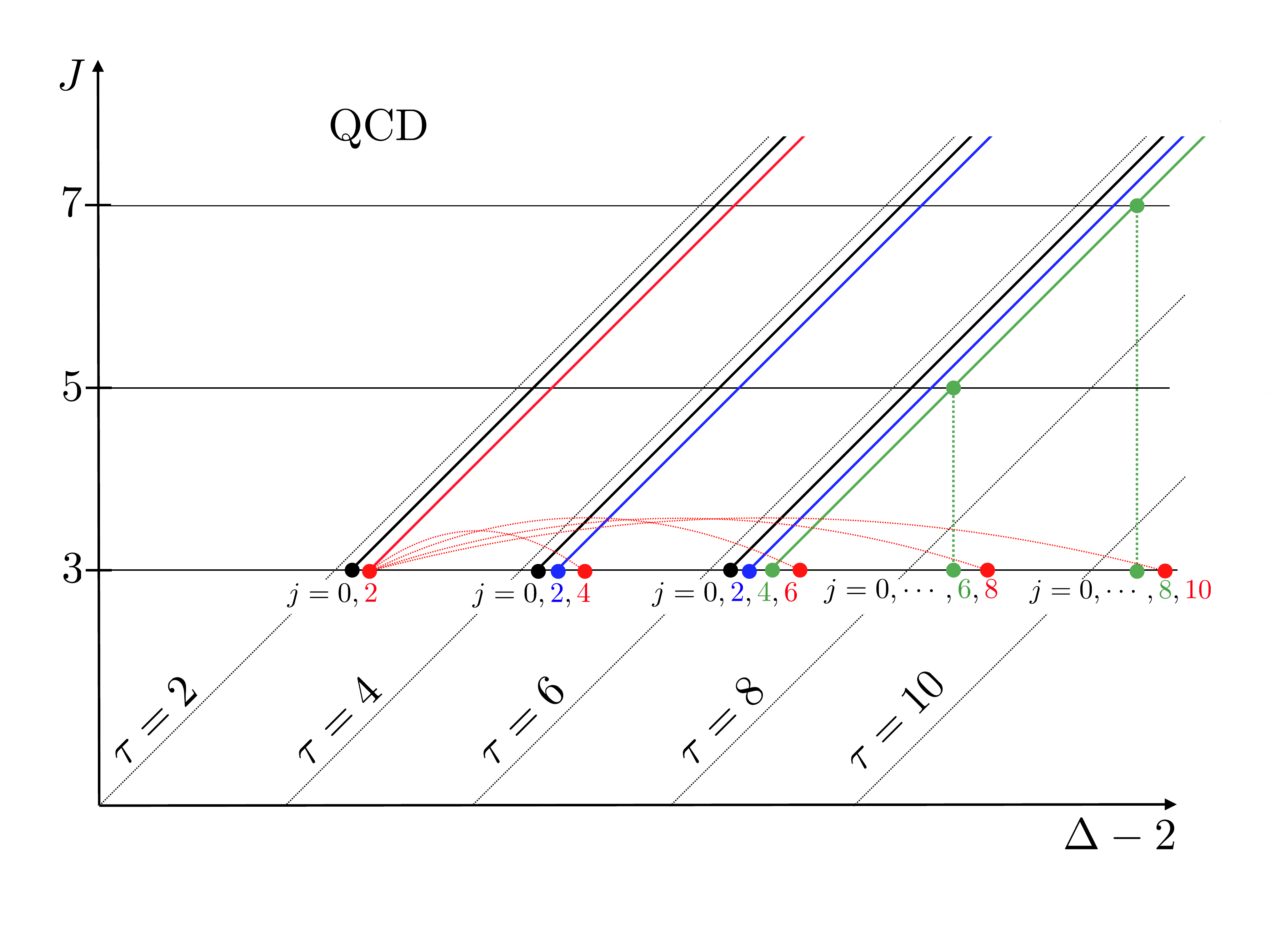}\label{fig:chewfrautschi_a}
}
\subfloat[]{
\includegraphics[scale=0.18]{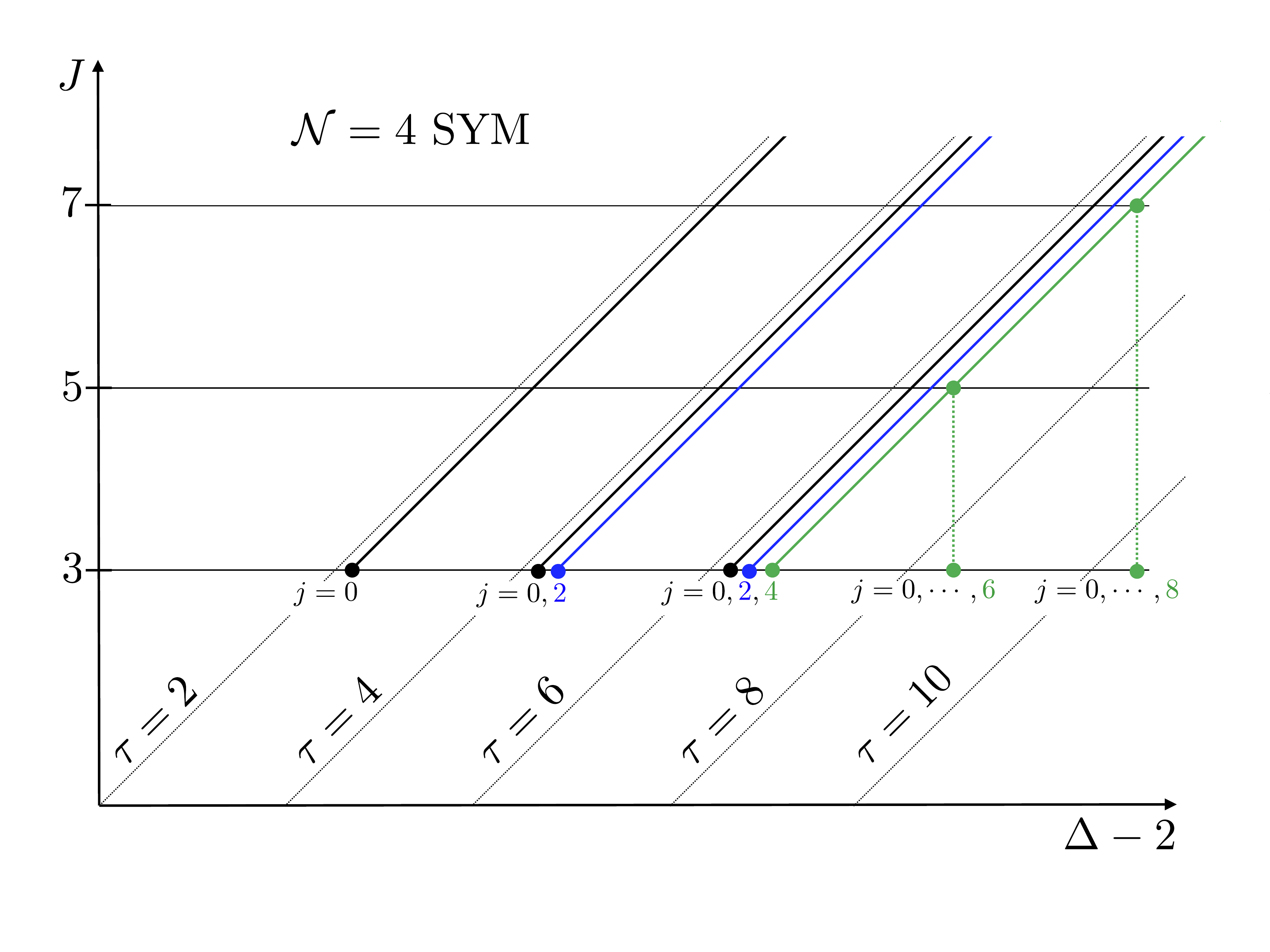}\label{fig:chewfrautschi_b}
}\qquad
\end{center}
\caption{Chew-Frautschi plot relevant for the squeezed limit of the three-point correlator in QCD, (a), and in $\cN=4$ SYM, (b). As compared to the structure in a generic field theory, at leading order in the perturbative expansion, one does not have contributions from $j=4$ operators at twist-2 or twist-4. This has the consequence that in QCD, the highest transverse spin terms at each twist are descendants of the twist-2 $j=2$ operator. In $\cN=4$, supersymmetry eliminates these contributions, leading to a simplified spectrum of operators. A more detailed discussion is presented in the text.}
\label{fig:chewfrautschi}
\end{figure}

We begin by discussing the structure of transverse spin terms in the OPE limit. Following \cite{1822249}, this is most easily done by studying the Chew-Frautschi plot for the operators contributing, which is shown in \Fig{fig:chewfrautschi_a} for the case of QCD. Due to the fact that our calculation is the leading order result in a perturbative gauge theory, it has a significantly simplified structure, as compared a generic field theory. We recall that the general formula for the OPE takes the form \cite{1822249}
\begin{align}\label{eq:EEOPE}
\cE(\vec n_1) \cE(\vec n_2) &=\sum\limits_i \big( \mathbb{O}_{i,J=3,j=0}(\vec n) + \mathbb{O}_{i,J=3,j=2}(\vec n)  + \mathbb{O}_{i,J=3,j=4}(\vec n)   \big)\nn \\
&+\sum\limits_{n,i} \cD_{2n} \mathbb{O}_{i,J=3+2n,j=4}(\vec n)   \,,
\end{align}
and we will attempt to frame our discussion in this language. Here $\mathbb{O}$ are lightray operators labelled by a spin-$J$ and a transverse spin-$j$, whose explicit form will be given for the leading twist operators in \Sec{sec:ope}, and $\cD_{2n}$ are differential operators \cite{Karateev:2017jgd,1822249} which convert $2n$ units of spin to transverse spin. These differential operators are responsible for the higher transverse spin primaries in our result. In this paper we will primarily focus on the structure of the highest transverse spin contributions. A complete analysis of the perturbative result in terms of the light-ray OPE will be presented in future work \cite{blocks:forthcoming}.

We begin by considering the leading twist ($\tau=2$) contributions, where we have operators with $j=0$ and $j=2$.
First, as was highlighted in \cite{Chen:2020adz}, in QCD there is a leading power, transverse spin-2 contribution, which is phenomenologically relevant. The resummation of the all orders logarithmically enhanced terms in the squeezed limit associated with this leading power transverse spin-2 contribution was performed in \cite{Chen:2020adz} using the QCD light-ray OPE, and will be discussed in more detail in this paper. This contribution is universal between quarks and gluons since it can only arise from gluonic twist-2 operators. In a weakly coupled gauge theory, there are no leading twist $j=4$ operators. This trajectory is therefore absent in \Fig{fig:chewfrautschi_a}, as are all contributions obtained by acting on it with the differential operators $\cD_{2n}$.

We now consider the extension to higher twists. Here it is important to emphasize that since we work to leading order in the perturbative expansion, we can only probe operators with up to 4 fields. Operators with additional fields cannot contribute at this order in perturbation theory, but could certainly contribute at higher orders. Furthermore, operators with four fields can only contribute through a direct contraction of the fields, placing further constraints on how they can contribute. Most importantly, there is a twist-4 transverse spin operator that takes the schematic form (see also the discussion in \cite{1822249})
\begin{align}\label{eq:twist4_op}
F \partial^{J_1}F \partial^{J_2}F\partial^{J_3}F\,,
\end{align}
with $J_1+J_2+J_3+4=J$. This operator only exists as a local operator when $J\geq 4$, but naively one would expect its light-ray continuation to $J=3$ to contribute to the OPE.\footnote{While there are some older studies of analyticity in spin of higher twist-3 and twist-4 operators in QCD (e.g. \cite{Ali:1991em,Belitsky:1997zw,Belitsky:1997by,Braun:2009vc,Ji:2014eta}), it seems interesting to revisit them with a modern perspective in light of recent advances in CFT. } However, when its indices are contracted to obtain its tree level contribution, the transverse spin-4 contribution from this operator vanishes. Therefore, there is no $j=4$ twist-4 operator contributing to our result, and correspondingly, no contributions from any of its $\cD_{2n}$ descendants. 

Because of these cancellations, at each power in the twist expansion, the highest transverse spin contributions arise only as descendants of the leading twist-2 $j=2$ operator. These are shown schematically in  \Fig{fig:chewfrautschi_a} by the red dashed lines connecting these contributions to the leading twist operator. This has two important consequences. First, the universality of the highest transverse spin contribution between quark and gluon jets holds to all powers in the twist expansion. Second, since these particular contributions are uncontaminated, it allows us to directly verify the structure of celestial blocks from our perturbative calculation.  For any given color structure, we find that the highest transverse spin contributions are proportional to 
\begin{align}
&\frac{1}{\bar{w}^2}\, _2F_1\left[1,  \frac{3}{2},\frac{7}{2},w^2 \right]+\frac{1}{w^2} \, _2F_1\left[1,  \frac{3}{2},\frac{7}{2},\bar w^2\right] \\
&\hspace{3cm}=\frac{\cos 2\phi}{|w|^2}+\frac{3}{7} \cos 4\phi  +|w|^2 \frac{5}{21}\cos6\phi +|w|^4 \frac{5}{33}\cos8\phi  +\cdots\,. \nn
\end{align}
This can be identified with a 2D conformal block on the plane transverse to the jet, indicating that by studying the collinear limit of celestial objects we can restore the conformal symmetry on the celestial sphere~\footnote{Which is simply the Lorentz symmetry $\text{SO}(3,1)$ in $3+1$ dimension.} which is generically broken by the source in the bulk. By dimensional analysis and symmetry, we can view the energy flow operator $\mathcal{E}$ as a scalar operator with celestial dimension 3 while the unpolarized quark source can fictitiously be regarded as a celestial dimension 5 scalar on the celestial sphere. Then the Casimir equation for conformal blocks corresponding to this collinear 3-point energy correlator are \cite{blocks:forthcoming}
\bea
2z\left((1-z)^2G_{h}^{\prime \prime}(z)-6(1-z)G_{h}^{\prime}(z)+6G_{h}(z)\right)=2h(h-1) G_{h}(z)\,,\\
2\bar{z}\left((1-\bar{z})^2G_{\bar{h}}^{\prime \prime}(\bar{z})-6(1-\bar{z})G_{\bar{h}}^{\prime}(\bar{z})+6G_{\bar{h}}(\bar{z})\right)=2\bar{h}(\bar{h}-1) G_{\bar{h}}(\bar{z})\,,
\eea
where we have decomposed the conformal block into holomorphic part and anti-holomorphic part $G_{\delta,j}(z,\bar{z})=G_{h}(z)G_{\bar{h}}(\bar{z})+G_{\bar{h}}(z)G_{h}(\bar{z})$. $\delta$ is the celestial dimension that relates to the bulk dimension $\Delta$ through $\delta=\Delta-1$ and $j$ is the transverse spin. $h=\frac{\delta-j}{2}$ and $\bar{h}=\frac{\delta+j}{2}$ are the standard conformal weights in 2D CFT. In terms of the $w$ variable, if we define $\widetilde{G}_{h}(w)=\frac{2}{(1+w)^2}G_{h}(z)$, the differential equation becomes
\beq
4(1-w^2)\left(2 w^4 \partial_{w^2}^2+7w^2 \partial_{w^2}+3 \right)\widetilde{G}_{h}(w)=2h(h-1)\widetilde{G}_{h}(w)\,,
\eeq
The solutions consistent with collinear twist expansion when $h=3,0$ are
\beq
\widetilde{G}_{h=3}(w)=\,_2F_1\left( 1, \frac{3}{2}, \frac{7}{2}, w^2\right)\quad \text{and}\quad \widetilde{G}_{h=0}(w)=\frac{1}{w^2}, \nonumber
\eeq
which agree exactly with our perturbative result.\footnote{Note that by changing the normalization of the correlator, and working in the $z,\bar z$ variables, these are the standard two dimensional conformal blocks.} The ability to use symmetries to resum subleading twist corrections into blocks is broken by many more standard jet substructure observables, and highlights the elegant structure of the energy correlators, especially in the collinear limit. In particular, the collinear limit of 3 point energy correlator is quite similar to a 4 point correlation function and  admits a similar conformal block decomposition. A detailed derivation of the blocks for generic quantum numbers, and a decomposition of the perturbative result into blocks will be presented in future work \cite{blocks:forthcoming}. The ability to isolate this block in the QCD result by choosing specific quantum numbers provides a strong check on the validity of the blocks.  The resummation of descendants of twist-2 operators in QCD was also recently considered in \cite{Braun:2020zjm}.

In the collinear limit, where the energy flow operators live in a two dimensional plane transverse to the jet, one expects a close connection to the Regge limit, where the non-trivial dynamics also lives in the two-dimensional plane transverse to the scattering. While it is well known that the anomalous dimensions that govern the scaling of the energy correlators can be analytically continued to the Regge limit, $J=1$, it would be extremely interesting to do this for the shape dependence of a higher point correlator, perhaps to make contact with 
\cite{Balitsky:2013npa,Balitsky:2015tca,Balitsky:2015oux}. We see this close connection at the level of the blocks, where we can interpret the block as the Gegenbauer Q-function
\begin{align}
Q_{J=3}^{(d=2)}=\left. \,_2F_1\left( \frac{J+d-3}{2},\frac{J+d-2}{2}, J+\frac{d-1}{2},\frac{1}{z^2}  \right)\right|_{J=3,d=2}\,,
\end{align}
familiar from the case of the partial wave expansion or the study of the Regge limit (see e.g. \cite{Correia:2020xtr} for a modern, comprehensive discussion). For this particular case, we have $J=3$ associated with the collinear spin of the squeezed operators, and $d=2$ associated with the celestial sphere. In the Regge limit of massless scattering, and for t-channel partial waves, one has $1/z^2=t/(2s)$, which in this case is associated with the variable $w$ in the squeezed limit. It would be interesting to study this connection in more detail.

Finally, we comment on the higher twist terms with $\log(2|\omega|)$ dependence, highlighted in magenta.
As seen in the integrand level expansion in \Sec{sec:expansion}, if we consider the OPE limit of detectors 1 and 2, in a perturbative gauge theory calculation, terms with dependence on $\log(2|\omega|)$ arise from the limit where the particle detected by the third detector is soft. Due to the energy weighting of the detector, such terms can first appear at subleading twist, in agreement with the explicit results given in \Eqs{eq:quark_result_expand}{eq:gluon_result_expand}. In this limit, the standard divergence associated with the emission of a soft gluon in a gauge theory manifests as a logarithm. These logarithmic terms seem closely related to so called ``endpoint divergences" in gauge theory factorization, due to the fact that in this limit the two squeezed partons carry all of the energy. Endpoint divergences occur ubiquitously in gauge theory factorization, particularly at subleading power. There has been significant recent work in understand subleading power factorization for double logarithmic Sudakov type observables (see e.g. \cite{Moult:2018jjd,Beneke:2018gvs,Moult:2019vou,Moult:2019uhz,Beneke:2019mua,Liu:2019oav,Liu:2020tzd,Liu:2020wbn}), and it would be interesting to precisely connect the two. For the particular case of the energy correlators, such terms should be reproduced by the light-ray OPE, and therefore a detailed understanding of the correspondence may help in the study of endpoint divergences more generally.\footnote{We thank Cyuan-Han Chang and David Simmons-Duffin for helpful discussions related to these terms.}

\subsection{$\cN=4$ SYM}

For the much simpler case of $\cN=4$ SYM, we obtain
\beq \label{eq:N=4_expansion}
\begin{split}
&\widetilde{G}_{\mathcal{N}=4}(w)=
\frac{1}{\left| w\right|^2}+\left[\left(\frac{44}{3} -\frac{4}{3} \pi ^2 \right)\cos (2 \phi )-\pi ^2+\frac{41}{3}\right]\\
&+\left| w\right| ^2\! \left[\! {\color{magenta}\left(\frac{32}{5}-\frac{8}{5}\cos (2 \phi )\right)\! \log (2\!  \left| w\right| )}
    +\left(\frac{416}{15} -\frac{8}{3} \pi ^2 \right)\! \cos (4 \phi )
   +\left(\frac{1688}{25} -6 \pi ^2\right)\! \cos (2 \phi)
  -\frac{4 \pi ^2}{3}+\frac{1159}{75}\right]\\
&+ \left| w\right| ^4 \left[{\color{magenta}\left(\frac{128}{7} \cos (2 \phi )-\frac{16}{7} \cos (4 \phi)+\frac{2776}{105}\right) \log (2 \left| w\right| )}\right.\\
&\left. +\left(\frac{64346}{1575}-4 \pi ^2\right)\! \cos (6 \phi )
+\left(\frac{393634}{3675}-10 \pi ^2\right)\! \cos (4 \phi )
+\left(\frac{37462}{735}-4 \pi ^2\right)\! \cos (2 \phi)
-5 \pi^2+\frac{571889}{11025}\right]\\
&+\cO(|w|^6)\,.
\end{split}
\eeq

As with the case of the QCD result, this result is best understood in terms of the Chew-Frautschi plot for the contributing operators which is illustrated in \Fig{fig:chewfrautschi_b}. The case of $\cN=4$ SYM is even more special than QCD, since in addition to the generic simplifications in a weakly coupled gauge theory described in  \Sec{sec:QCD_power_discussion}, in $\cN=4$ SYM the contribution from the leading twist $j=2$ operators cancel due to supersymmetry \cite{Chen:2020adz}. We will explain the origin of this cancellation in more detail in  \Sec{sec:amp}. Since the highest transverse spin contributions at each twist arise as descendants of this leading twist operator, this implies that the highest transverse spin contributions vanish in $\cN=4$ SYM at each power in the twist expansion.  This gives rise to a  ``leading twist classicality" in $\cN=4$, as was noted in \cite{Chen:2020adz}. 

As can be seen from \Fig{fig:chewfrautschi_b}, this large degree of simplification implies that the ``generic" expected structure in the squeezed limit does not start until quite high twists. In particular, since transverse spin-4 contributions first appear at twist-6 ($\cO(|\omega|^2)$), contributions from $\cD_{2n}$ operator acting on this Regge trajectory start at twist-8 ($\cO(|\omega|^4)$) and transverse spin-6. Therefore it seems that supersymmetry plays a significant role in suppressing the contributions from higher transverse spin contributions, making them harder to study. In this respect, the QCD result, although more complicated, is perhaps more representative of the generic structure that one expects. A complete analysis of this result at subleading twists will be presented in future work \cite{blocks:forthcoming}. While there is some understanding of the twist operators in $\cN=4$ \cite{Belitsky:2003sh,Belitsky:2004sc,Belitsky:2004yg,Belitsky:2005gr}, the structure of such high twists is not known.

In \Eq{eq:N=4_expansion} we have again highlighted the logarithmically enhanced terms in magenta, which illustrates that they are not specific to QCD, and are therefore not related to the breaking of conformal symmetry.  We hope that the perturbative data provided in this section from the full calculation of the three-point correlator will be useful for the further development of the light-ray OPE.

\section{Leading Twist Calculation from Splitting Amplitudes}\label{sec:amp}

Although our ultimate goal is to understand the structure of the squeezed limit using the light-ray OPE, we begin by re-calculating the leading power behavior using a standard splitting function calculation. This calculation is simple, but serves two purposes. First, it provides an independent cross check of the result for the full three-point correlator presented in \cite{Chen:2019bpb}. Second, it allows us to describe the origin of the transverse spin structure in the squeezed limit in a weakly coupled gauge theory from a particle physics perspective. In particular, we will see that it arises from the interference of intermediate gluons with different helicities, as was described in  \cite{Chen:2020adz}.

For later convenience, we rewrite the leading power term in QCD as\footnote{Here we used the relation $\frac{d^3\Sigma_i}{d\theta_L^2 d\theta_S^2 d \phi}=\frac{1}{32 \theta_L^2} \frac{d^3\Sigma_i}{d x \; d\mathrm{Re} w\; d\mathrm{Im} w}\bigg\vert_{x\to\frac{\theta_L^2}{4}, \abs{w}\to \frac{\theta_S}{2\theta_L}, \arg{w}\to \phi}$. At leading power, we also have $\frac{1}{\sigma_{\rm tot}} \frac{d^3 \Sigma}{d\theta_L^2\, d\theta_S^2\, d\phi} = \pi \langle \mathcal{E}(\vec{n}_1) \mathcal{E}(\vec{n}_2) \mathcal{E}(\vec{n}_3) \rangle$.}
\beq
 \frac{d^3 \Sigma_i}{d\theta_L^2 d\theta_S^2 d\phi} \simeq  \frac{1}{\pi}  \left(\frac{\alpha_s}{4 \pi} \right)^2   \frac{ \text{Sq}_i^{(0)}(\phi)}{\theta_L^2 \theta_S^2} +\cdots\,,
\eeq
where
\bea
 \text{Sq}^{(0)}_q (\phi)&=&C_F n_f T_F \left(  \frac{39-{\color{blue}20\cos(2\phi)}}{225} \right) 
                 +C_F C_A \left(  \frac{273+{\color{red}10\cos(2\phi)}}{225} \right)+C_F^2 \frac{16}{5}\,, \label{eq:squeezed_quark}\\
\text{Sq}^{(0)}_g (\phi) &=&C_A n_f T_F  \left(  \frac{126-{\color{blue}20\cos(2\phi)}}{225} \right)
+C_A^2 \left(  \frac{882+{\color{red}10\cos(2\phi)}}{225} \right)+C_Fn_f T_F \frac{3}{5}\,. \label{eq:squeezed_gluon}
\eea
Here we have highlighted in blue the azimuthal dependence from a quark anti-quark pair going through the squeezed detectors, and in red the azimuthal dependence from a pair of gluons going through the squeezed detectors. These two terms have opposite signs due to the opposite statistics of the particles, leading to a partial cancellation. In $\mathcal{N}=4$ SYM, when the contribution from scalars is added, this cancellation is exact and the $\cos(2\phi)$ dependence is absent at leading twist, as seen in \Eq{eq:N=4_expansion}. Since we will interpret the $\cos(2\phi)$ as a quantum interference effect, this cancellation adds to the well known classical nature of $\mathcal{N}=4$ SYM theory. Finally, the third term in \Eqs{eq:squeezed_quark}{eq:squeezed_gluon} arises from a quark-gluon pair going through the squeezed detectors. This term has no azimuthal dependence due to the the fact that massless quark helicity is conserved.

\begin{figure}[htbp]
\begin{center}
\includegraphics[width=5cm]{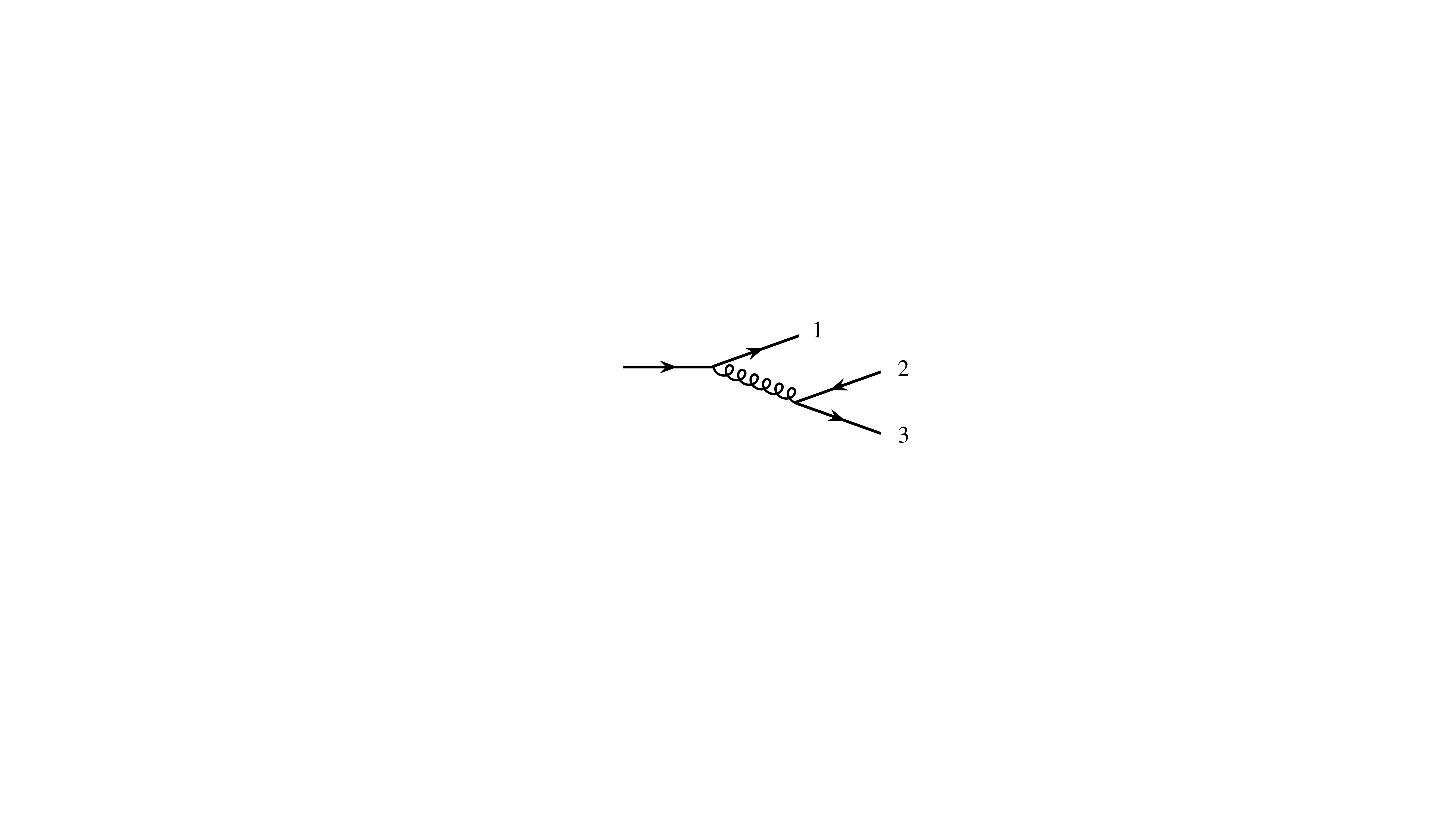}
\caption{Calculation of the $n_f$ contribution in the squeezed limit using iterated splitting functions. The non-trivial azimuthal dependence arises from the interference between the splittings of gluons of different helicities.}
\label{splitting_diagram}
\end{center}
\end{figure}

To illustrate the interpretation of the $\cos(2\phi)$ transverse spin term from a standard splitting function calculation, we choose for simplicity the $n_f$ term in  \Eq{eq:squeezed_quark}. The relevant Feynman diagram is shown in \Fig{splitting_diagram} where particles $2$ and $3$ denote the squeezed pair. At leading twist, this diagram factorizes into a product of iterated $1\to 2$ splittings. Suppose the helicity of the incoming quark source is $+$ (is $-$ in all outgoing convention), then the outgoing quark $1$ has $+$ helicity due to helicity conservation. For the same reason, the quark anti-quark pair $2,3$ also have opposite helicity. Then using the factorization in squeezed limit, we obtain
\beq
\begin{split}
\mathcal{M}(q^+\to 1^+ + 2^+ + 3^-)&=\mathrm{Split}_{-}^{q\to q_1 g}(1^+, P^+) \times \mathrm{Split}_{-}^{g\to q_2^\prime \bar{q}_3^\prime}(2^+,3^-)\\
&+ \mathrm{Split}_{-}^{q\to q_1 g}(1^+, P^-) \times \mathrm{Split}_{+}^{g\to q_2^\prime \bar{q}_3^\prime}(2^+,3^-)\,,
\end{split}
\eeq
where $P$ denotes the momentum of the virtual gluon whose helicity could be $+$ or $-$ and therefore could interference with each other. All relevant splitting functions can be found in \cite{Bern:1999ry}\footnote{The conventions for splitting amplitudes used in this paper iss $\sqrt{2}$ times that in \cite{Bern:1999ry}. }  
and with a little manipulation we find
\beq
\begin{split}
\mathcal{M}(q^+\to 1^+ + 2^+ + 3^-)&= -\frac{\xi_2}{(\xi_2+\xi_3)^{\frac{3}{2}}}\frac{2}{\langle1\, P\rangle \left[2\,3\right]}+ \frac{\xi_1 \xi_3}{(\xi_2+\xi_3)^{\frac{3}{2}}}\frac{2}{\left[1\, P\right] \langle 2\,3\rangle}\\
&= \frac{2}{\theta_L \theta_S}\left(\frac{\sqrt{\xi_2/(\xi_1 \xi_3)}}{(\xi_2+\xi_3)^2}e^{-i(\phi_L-\phi_S)} - \frac{\sqrt{(\xi_1 \xi_3)/\xi_1}}{(\xi_2+\xi_3)^2}e^{i(\phi_L-\phi_S)}\right)\,,
\end{split}
\eeq
where $\xi_i$ is the momentum fraction relative to the incoming source. We can separate the amplitude squared into a non-interference and an interference piece and find that the $\cos(2\phi)$ is an interference effect in spin space:
\bea
\text{non-interference: }&\quad \frac{4}{\theta_L^2\theta_S^2}\frac{\xi_2^2 +\xi_1^2 \xi_3^2}{\xi_1\xi_2\xi_3 (\xi_2+\xi_3)^4}\,,\\
\text{interference: }&\quad  -\frac{4}{\theta_L^2\theta_S^2}\frac{2\cos\left(2\phi\right)}{(\xi_2+\xi_3)^4}\,.
\eea

The phase space and energy weighting to obtain $\displaystyle \frac{d^3\Sigma_i}{d\theta_L^2 d\theta_S^2 d\phi}$ is
\[
  \displaystyle \frac{1}{\pi} \frac{g^4}{(4\pi)^4}\int\! d\xi_1 d\xi_2 d\xi_3 \delta(1-\xi_1-\xi_2-\xi_3) \xi_1^2 \xi_2^2 \xi_3^2
  \, .\]
In addition, the exchange of helicity or particle species for the squeezed pair give the same contribution, so we need to multiply $4$ to reproduce the $C_F n_f T_F$ term in quark jet result.

\begin{table}[htp]
\caption{Splitting amplitudes for gluons, fermions and scalars, and their associated interference terms. In $\cN=4$ SYM, the appropriately weighted sum over the supermultiplet leads to a vanishing spin interference effect at leading twist.}
\begin{center}
\begin{tabular}{|c|c|c|c|}
\hline
Splitting Pair  & $\left(G_+,  G_-\right)$ & $ \left( \Gamma_A, \overline{\Gamma}_A\right)$ & $\left( S_{AB}, S_{CD}\right)$\\ \hline
$\mathrm{Split}\left(G_{+} \to a+b \right)$ & $-\frac{z^2}{\sqrt{z(1-z)}\left[1\, 2\right]}$ & $\pm\frac{z^{3/2} (1-z)^{1/2}}{\sqrt{z(1-z)}\left[1\, 2\right]}$ & $\pm \frac{z(1-z)}{\sqrt{z(1-z)}\left[1\, 2\right]}$ \\ \hline
$\left[\mathrm{Split}\left(G_{-} \to a+b \right)\right]^{*}$ &   $-\frac{(1-z)^2}{\sqrt{z(1-z)}\left[1\, 2\right]}$   &   $\mp \frac{z^{1/2} (1-z)^{3/2}}{\sqrt{z(1-z)}\left[1\, 2\right]}$    &    $\pm \frac{z(1-z)}{\sqrt{z(1-z)}\left[1\, 2\right]}$    \\ \hline
Interference Term & $\frac{z(1-z)}{\left[1\, 2\right]^2} +\text{c.c.}$ & $-\frac{z(1-z)}{\left[1\,2\right]^2}+\text{c.c.}$ & $\frac{z(1-z)}{\left[1\,2\right]^2}+\text{c.c.}$ \\ \hline
\end{tabular}
\end{center}
\label{table:split_N=4}
\end{table}%

The procedure for $\mathcal{N}=4$ SYM theory is exactly the same, so we will omit the calculation but show how interference effect is cancelled. Interference only happens when the virtual gluon splits into a pair with total helicity $0$. Through the relevant splitting amplitudes listed in Table.\ref{table:split_N=4}, the cancellation happens in $\mathcal{N}=4$ SYM since there are 2 gluons, 8 gluinos and 6 gluons and the existence of relative minus sign between different statistics.

Therefore, we see that from a standard particle physics perspective, the transverse spin effects in the squeezed limit arise from the interference between different helicities of the intermediate gluons in the jet. This perspective was advocated in \cite{Chen:2020adz}, where it was shown that studying this squeezed limit in jet substructure at the LHC provides an interesting window into the spin behavior of the gluon, and quantum interference effects.

\section{The Light-Ray OPE in QCD}\label{sec:ope}

Having presented the structure of the twist expansion in the OPE limit using our perturbative results from \cite{Chen:2019bpb}, and reproduced the leading power fixed order result using a standard splitting function calculation, in this section we develop the light-ray OPE in QCD with a focus on understanding the all orders perturbative structure of the leading twist OPE limit, using the renormalization group evolution of the light-ray operators appearing in the OPE.

To understand the squeezed limit of the three-point correlator requires an iterated light-ray OPE. We can first perform the OPE $\mathcal{E}(\hat n_2) \mathcal{E}(\hat n_3) \to \sum \mathbb{O}^{[3]}(\hat n_2)$, followed by $\mathcal{E}(\hat n_1)  \vec{\mathbb{O}}^{[J]}(\hat n_2)  \to \vec{\mathbb{O}}^{[J+1]} (\hat n_1)$. We will see that these OPEs package the transverse spin quantum numbers in a convenient way. At leading twist, in a weakly coupled gauge theory, this iterated OPE closes onto a finite set of operators $\vec{\mathbb{O}}^{[J]}(\hat n)$ with spin $J$, and transverse spin $j=0,2$. We therefore perform all our calculations in this section for generic spin-$J$ so that our results can be used for arbitrary iterated limits of energy flow operators.

\subsection{Review of Light-Ray Operators}

Light-ray operators have long been studied in the QCD literature \cite{Balitsky:1987bk,Balitsky:1988fi,Balitsky:1990ck,Braun:2003rp}, including more recently in the BFKL limit \cite{Balitsky:2013npa,Balitsky:2015oux,Balitsky:2015tca}. Their use in event shapes was first introduced in \cite{Hofman:2008ar}, where it was argued that one could perform an OPE of energy flow operators, and that their small angle asymptotics were governed by the anomalous dimensions of twist-2 spin-3 operators.  More recently, in the context of CFTs, light-ray operators have been studied more formally \cite{Kravchuk:2018htv}, and argued to give the natural analytic continuation in spin \cite{Caron-Huot:2017vep} of local CFT operators. In this context, the OPE between light-ray operators has been developed into a rigorous OPE \cite{Kologlu:2019mfz,Kologlu:2019bco,1822249}. We note that analyticity in spin at the level of explicit anomalous dimension has been applied to make connection between DGLAP and BFKL limit of ${\cal N}=4$ SYM~\cite{Kotikov:2002ab,Kotikov:2003fb,Kotikov:2004er,Kotikov:2005gr}. Very recently it has also motivated the experimental implementation in terms of jet substructure~\cite{Chen:2020vvp}.

While the intuition for the existence of the OPE does not rely on conformal symmetry, in the absence of conformal symmetry, one has significantly less control over the structure of the OPE. Therefore, in this paper, we will focus on understanding its structure in perturbation theory and at leading twist, which is already sufficient for many phenomenological applications.

We now briefly review the notion of light-ray operators\footnote{In this paper, we call the object defined on a light-ray as light-ray operators and thus do not distinguish the terminology light transform of local operator and light-ray operator, while in the discussion of \cite{Kravchuk:2018htv}, the light transform of local operator can roughly be regard as the residue of light-ray operator located at corresponding poles.} and their basic properties. For simplicity, we restrict ourselves to the case of light-ray operators associated to local operators, as that is sufficient for our purposes. We refer the reader to \cite{Kravchuk:2018htv} for a comprehensive discussion. Given a local dimension-$\Delta$, spin-$J$, transverse spin-$j$ operator $\mathcal{O}^{\mu_1\dots\mu_J;\nu_1\dots\nu_j}$, we can define a corresponding nonlocal operator $\mathbb{O}(\vec{n})$ on a light-ray in a similar manner to the definition of energy flow operator
\beq
\mathbb{O}(\vec{n},\epsilon)=\lim_{r\to \infty} r^{\Delta-J}\int_{0}^{\infty}\! dt\; \mathcal{O}^{\mu_1\dots\mu_J;\nu_1\dots\nu_j}(t,r\vec{n})\bar{n}_{\mu_1}\dots\bar{n}_{\mu_J}\epsilon_{\nu_1}\dots\epsilon_{\nu_j}
\eeq
where $\bar{n}=(1,-\vec{n})$ is the direction of the light-ray located at future null infinity and $\epsilon$ is a polarization vector that satisfies $\epsilon^2=\epsilon\cdot n = 0$. Roughly speaking, in a perturbative picture, such an operator represents detecting single or multiple massless particles that are moving along the null direction $n$. With this construction, we can see the dimension of $\mathbb{O}(\vec{n},\epsilon)$ is $J-1$.

We may also regard $\mathbb{O}(\vec{n},\epsilon)$ as a local operator on the celestial sphere so that the 4D Lorentz group acts as the 2D conformal group with respect to its coordinate $\vec{n}$. For simplicity, we can discuss the little group that fixes $\vec{n}$ on the celestial sphere, which consists of boosts along $\vec{n}$ (corresponding to dilatations on the celestial sphere) and rotations around the axis $\vec{n}$ (corresponding to rotations on the celestial sphere). The boost quantum number (collinear spin) of $\mathbb{O}(\vec{n},\epsilon)$ is $1-\Delta$ (therefore the eigenvalue of dilatation on the celestial sphere is $\Delta-1$) because the transformation $\displaystyle \lim_{r\to\infty} r^{\Delta-J}\int_{0}^{\infty}\! dt$, which is proportional to $\displaystyle \lim_{x^{+}\to 0}(x^+)^{\Delta-J}\int_{-\infty}^{\infty}dx^{-}$, contributes $-(\Delta-J-1)$ in addition to the  $-J$ from the local operator. It is also obvious that the transverse spin structure is not changed under such a transformation. Therefore, we can conclude $\mathbb{O}(\vec{n},\epsilon)$ transforms as a local operator on the celestial sphere with celestial dimension $\delta=\Delta-1$ and transverse spin $j$.

Throughout this section we will follow the conventions for the mode expansion of \cite{Belitsky:2013xxa,Belitsky:2013bja}. In perturbation theory, we assume that the in- and out-states in the definition of S-matrix are asymptotic Fock states of a free theory. Therefore, the light-ray operators we are interested in can be expressed with free theory creation and annihilation operators when acting on those Fock states. The mode expansion for the energy flow operator can be found in \cite{Belitsky:2013xxa,Belitsky:2013bja} and we apply the same procedure to QCD twist-2 operators
\bea
\mathcal{O}_q^{[J]}&=&\frac{1}{2^J}\bar{\psi}\gamma^{+}(iD^+)^{J-1}\psi=\frac{1}{2^J} \bar{\psi}\gamma^+ (i\partial^+)^{J-1}\psi+\cdots,\\
\mathcal{O}_{g}^{[J], ij}&=&-\frac{1}{2^J} F_{c}^{(i +}(iD^+)^{J-2}F_{c}^{j) +}=-\frac{1}{2^J} \left(\partial^+ A^{i}_{c}\right)(i\partial^+)^{J-2}\left(\partial^+ A_c^{j} \right)+\cdots\,,
\eea
where $i,j$ are the transverse indices and $\cdots$ represents terms that are higher order in gauge coupling or vanish after some projections given below.

With the following convention for the mode expansions of massless free quark and free gluon field,
\bea
\psi(x)&=&\sum_s \int\frac{dp^+ d^2p_{\perp}}{(2\pi)^3 2p^+} \left(u_s(p) b_{p,s} e^{-i p\cdot x} + v_s(p) d^{\dagger}_{p,s} e^{i p\cdot x}\right)\,,\\
A^{\mu}_{c}(x)&=&\sum_{\lambda} \int\frac{dp^+ d^2p_{\perp}}{(2\pi)^3 2p^+}\left(\epsilon^{\mu}_{\lambda}(p) a_{p,\lambda,c} e^{-ip\cdot x} +{\epsilon^{*}_{\lambda}}^{\mu}(p) a^{\dagger}_{p,\lambda,c} e^{ip\cdot x} \right)\,,
\eea
we can obtain the corresponding light-ray operators
\bea
\hspace{-1cm}\mathbb{O}_q^{[J]}(\vec{n})&=&\sum_{s}\int\!\! \frac{E^2 dE}{(2\pi)^3 2E}  E^{J-1} \left(b_{p, s}^{\dagger}b_{p,s}+(-1)^J d_{p,s}^{\dagger} d_{p,s} \right)+\cdots\,, \label{eq: unpolarized_quark}\\
\hspace{-1cm}\mathbb{O}_g^{[J],ij}(\vec{n})&=-\frac{1}{2}&\sum_{\lambda,\lambda^\prime} \int\!\! \frac{E^2 dE}{(2\pi)^3 2E}  E^{J-1} \left( {\epsilon^{*}_{\lambda^\prime}}^{i}(p) \epsilon_{\lambda}^{j}(p)  +(-1)^{J} \epsilon_{\lambda}^{i}(p){\epsilon_{\lambda^\prime}^{*}}^{j}(p) \right) a^{\dagger}_{p,\lambda^\prime,c} a_{p,\lambda,c}+\cdots\,,
\eea
where $p^\mu=(E, E\vec{n})$ is the on-shell momentum and normal ordering is implicitly used. The $(-1)^J$ indicates we should analytic continue the $J$ separately for odd and even collinear spin. In this paper, only the analytic continuation from the even spin branch is relevant for us because the  energy measurement does not distinguish particles and anti-particles. 

The gluonic twist-2 operator can be further decomposed into a scalar and a traceless symmetric operator in the transverse plane. The most natural way to separate the transverse spin-0 and spin-2 operator by the index-free notation,
\bea
\mathbb{O}_{g}^{[J]}(\vec{n})&=&g^{\perp}_{ij} \mathbb{O}_g^{[J],ij}(\vec{n})=\sum_{\lambda, c} \int\!\! \frac{E^2 dE}{(2\pi)^3 2E} E^{J-1} a^{\dagger}_{p,\lambda,c} a_{p,\lambda,c}+\cdots\,, \label{eq: unpolarized_gluon}\\
\mathbb{O}_{\tilde{g},\lambda}^{[J]}(\vec{n})&=&\epsilon_{\lambda, i}(p)\epsilon_{\lambda,j}(p)\mathbb{O}_g^{[J],ij}(\vec{n})=-\sum_{\lambda} \int\!\! \frac{E^2 dE}{(2\pi)^3 2E}  E^{J-1} a^{\dagger}_{p,\lambda,c} a_{p,-\lambda,c}+\cdots\,, \label{eq: polarized_gluon}
\eea
where $\sum_i \epsilon_{\lambda, i} \epsilon_{\lambda,i} = 0$ with $i$ summed over transverse indices.

\subsection{The Light-Ray OPE}

In this section we compute the $\mathcal{E}(\hat n_1) \mathcal{E}(\hat n_2)$, $\mathcal{E}(\hat n_1) \mathbb{O}^{[J]}(\hat n_2) $ and $\mathbb{O}^{[J_1]}(\hat n_1) \mathbb{O}^{[J_2]}(\hat n_2) $ OPEs at leading twist. Since this set of operators closes under iterated OPEs these results allow for the description of arbitrary squeezed limits of multipoint correlators in QCD. In particular, they will allow us to analyze the squeezed limit of the three-point correlator, including its transverse spin structure, using an iterated OPE.

We will use a standard perturbative matching calculation to derive the leading twist OPE. From (\ref{eq: unpolarized_quark}), (\ref{eq: unpolarized_gluon}) and (\ref{eq: polarized_gluon}), we see that to determine the OPE coefficients at leading twist, it is sufficient to perform our matching calculations using 1-particle states. In particular, we can obtain the coefficient for $\mathbb{O}_q^{[J]}$ by the relation
\beq
\langle 0| \psi(x) \, \mathbb{O}_q^{[J]}(\vec{n})\, \bar{\psi}(0) | 0 \rangle=\int \!\! \frac{E^2 dE}{(2\pi)^3 2 E} e^{-i E n\cdot x} E^J \slashed{n} + \cdots\,,
\eeq
and for $\mathbb{O}_g^{[J]}(\vec{n}),\, \mathbb{O}_{\tilde{g},\lambda}^{[J]}(\vec{n})$ through 
\bea
\langle 0 | A^{\nu}_{b} (x)\, \mathbb{O}_g^{[J]}(\vec{n}) \, A^{\mu}_a | 0\rangle &=& \delta_{ab} \int\!\! \frac{E^2 dE}{(2\pi)^3 2E} e^{-i E n\cdot x} E^{J-1}\left( \epsilon^{\mu}_{+} \epsilon^{\nu}_{-} + \epsilon^{\nu}_{+} \epsilon^{\mu}_{-}\right)+\cdots\,,\\
\langle 0 | A^{\nu}_{b} (x) \mathbb{O}_{\tilde{g}, \lambda}^{[J]}(\vec{n}) A^{\mu}_a(0) |0\rangle &=& -\delta_{ab} \int\!\! \frac{E^2 dE}{(2\pi)^3 2E}e^{-i E n\cdot x} E^{J-1} \epsilon^{\mu}_{\lambda} \epsilon^{\nu}_\lambda+\cdots\,,
\eea
where the polarization vectors can be conveniently expressed in terms of spinor helicity variables if we introduce the reference spinor labelled by $q$ with $k \cdot q \neq 0$,
\beq
\epsilon_{+}^{\mu}(k) =\frac{1}{\sqrt{2}} \frac{\langle k | \gamma^{\mu} | q ] }{\left[ q\, k \right]},\quad \epsilon^{\mu}_{-} (k)=\frac{1}{\sqrt{2}} \frac{\langle q | \gamma^{\mu} | k ] }{\langle k \, q\rangle}.
\eeq
Using this perturbative matching approach we compute the $\mathcal{E}(\hat n_1) \mathcal{E}(\hat n_2)$ OPE in \Sec{sec:EE_OPE}, the $\mathcal{E}(\hat n_1) \mathbb{O}^{[J]}(\hat n_2) $ OPE in \Sec{sec:OE_OPE} and the $\mathbb{O}^{[J_1]}(\hat n_1) \mathbb{O}^{[J_2]}(\hat n_2) $ OPE in \Sec{sec:OO_OPE}. 

\subsubsection{$\mathcal{E}(\vec{n}_1)\mathcal{E}(\vec{n}_2)$ OPE}\label{sec:EE_OPE}

\begin{figure}[htbp]
\begin{center}
\includegraphics[width=10cm]{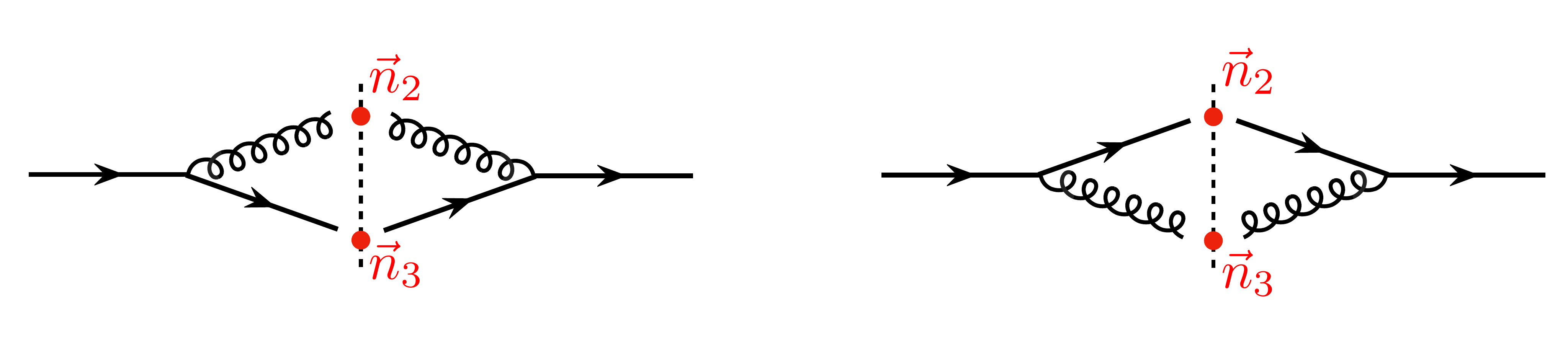}
\caption{Feynman diagrams contributing to the leading order matching of the $\mathcal{E}(\vec{n}_2)\mathcal{E}(\vec{n}_3)$ OPE onto the twist-2 quark operator $\mathbb{O}_{q}$.}
\label{fig: EE_OPE_1}
\end{center}
\end{figure}

The first Wightman function to calculate is $\langle 0 |  \psi(x) \mathcal{E}(\vec{n}_2)\mathcal{E}(\vec{n}_3) \bar{\psi}(0) | 0\rangle$ in which we omit the Wilson lines since we work with physical polarization and higher order corrections in the gauge coupling are beyond the scope of this paper. The source quark can split into a quark and a gluon that are respectively detected by two calorimeters located at $\vec{n}_2, \vec{n}_3$ (Figure \ref{fig: EE_OPE_1}). 

From the definition of the energy correlator and Feynman rules, we can obtain the LO result
\beq
g^2 C_F\int\!\! \frac{E_2^2 dE_2}{(2\pi)^3 2E_2} \frac{E_3^2 dE_3}{(2\pi)^3 2E_3} e^{-i (p_2+p_3)\cdot x} E_2 E_3\left(\sum_{\lambda} \frac{\slashed{p}_2+\slashed{p}_3}{(p_2+p_3)^2} \slashed{\epsilon}_\lambda(p_3) \slashed{p}_2 \slashed{\epsilon}_{-\lambda}(p_3) \frac{\slashed{p}_2+\slashed{p}_3}{(p_2+p_3)^2} + (2\leftrightarrow 3)\right)\,, \nonumber
\eeq
where the first term in the parenthesis can be written as
\beq
\frac{1}{2p_2\cdot p_3}\frac{2}{ [q\, 3] \langle 3\, q\rangle}\left( [q\,2]\langle 2\, q\rangle \slashed{p}_2 +([q\, 2]\langle 2\, q\rangle+ [q\, 3]\langle 3\, q\rangle)(\slashed{p}_2+\slashed{p}_3) -2p_2\cdot p_3 \slashed{q}  \right)\,. \nonumber
\eeq
In the collinear limit, $2p_2\cdot p_3\approx E_2 E_3 \theta_S^2,\, p_2^\mu\approx E_2 n^\mu,\, p_3^\mu\approx E_3 n^\mu $, the leading power contribution is 
\bea\label{eq:change_var}
& &\langle 0 |  \psi(x) \mathcal{E}(\vec{n}_2)\mathcal{E}(\vec{n}_3) \bar{\psi}(0) | 0\rangle\nn \\
&\approx& \frac{2}{\theta_S^2} g^2 C_F\!\! \int\! \frac{E_2^2 dE_2}{(2\pi)^3 2E_2} \frac{E_3^2 dE_3}{(2\pi)^3 2E_3}  \slashed{n}  \left(\frac{E_2^2+(E_2+E_3)^2}{E_3} +(2\leftrightarrow 3) \right) e^{-i (E_2+E_3)n\cdot x}\nonumber\\
&=& \frac{1}{2\pi}\frac{4}{\theta_S^2} \frac{g^2}{(4\pi)^2} C_F\!\! \int_0^1 \!\! dz\, z(1-z)\left(\frac{1+z^2}{1-z}+\frac{1+(1-z)^2}{z} \right) \int \!\! \frac{E^2 dE}{(2\pi)^3 2E} \slashed{n} E^3 e^{-i E n\cdot x}\,, \nonumber
\eea
where we have changed the variables $E_2=z E, \, E_3=(1-z) E$ to get the result in the second line. One can immediately identify the second integral as $\langle 0| \psi(x) \, \mathbb{O}_q^{[3]}(\vec{n})\, \bar{\psi}(0) | 0 \rangle$ which also confirms that only collinear spin-3 operators appear in the OPE when the theory is scale invariant \cite{Hofman:2008ar}. Together with the color factor $C_F$, the two terms in the parenthesis of the first integral are  the regular part of Altarelli-Parisi splitting kernels $P_{q\leftarrow q}(z)$ and $P_{g\leftarrow q}(z)$. The Mellin transform of the splitting kernels are related to the standard twist-2 anomalous dimensions through
\beq
\gamma_{ab} (J)=-2\int_0^1 dz\, z^{J-1} P_{a \leftarrow b}(z).
\eeq
With this convention, we obtain
\beq \label{eq: EE_OPE_quark_state}
\langle 0 |  \psi(x) \mathcal{E}(\vec{n}_2)\mathcal{E}(\vec{n}_3) \bar{\psi}(0) | 0\rangle
=-\frac{1}{2\pi} \frac{2}{\theta_S^2} \left[(\gamma_{qq}(2)-\gamma_{qq}(3))+(\gamma_{gq}(2)-\gamma_{gq}(3)) \right] \langle 0| \psi(x) \, \mathbb{O}_q^{[3]}(\vec{n})\, \bar{\psi}(0) | 0 \rangle+\mathcal{O}(\theta_S^0)\,.
\eeq

\begin{figure}[htbp]
\begin{center}
\includegraphics[width=10cm]{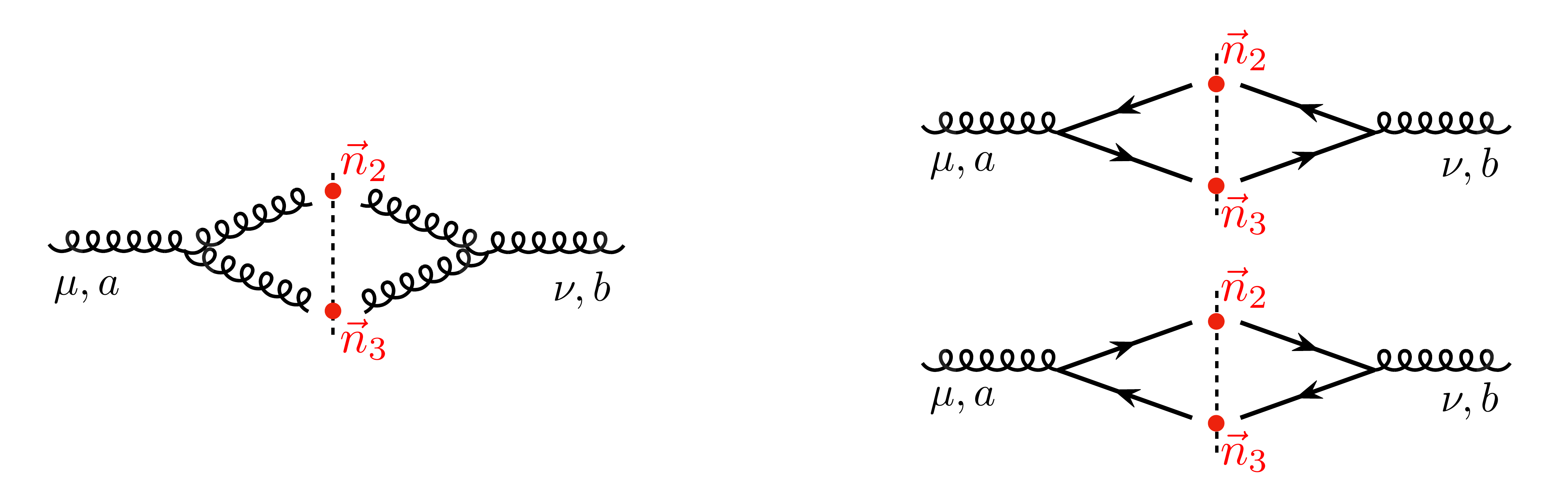}
\caption{Feynman diagrams contributing to the leading order matching of the $\mathcal{E}(\vec{n}_2)\mathcal{E}(\vec{n}_3)$ OPE onto the twist-2 gluon operator $\mathbb{O}_{g}$ and $\mathbb{O}_{\tilde{g},\lambda}$.}
\label{fig: EE_OPE_2}
\end{center}
\end{figure}

Similarly, we can perform the matching calculation for a gluon state (Figure \ref{fig: EE_OPE_2}). A gluon can either split into a pair of gluons 
\beq
\begin{split}
 g^2 C_A \frac{2}{(2p_2\cdot p_3)^2} &\left(  2\left(\frac{E_2}{E_3}+\frac{E_3}{E_2}+\frac{E_2 E_3}{(E_2+E_3)^2} \right)  \langle 2\, 3\rangle [3\,2] (\epsilon^{\mu}_{+}(n)\epsilon^{\nu}_{-}(n)+\epsilon^{\nu}_{+}(n)\epsilon^{\mu}_{-}(n))  \right. \\
 &\quad \left. +\frac{2E_2 E_3}{(E_2+E_3)^2}\left( [3\,2]^2 \epsilon^{\mu}_{+}(n)\epsilon^{\nu}_{+}(n)+\langle 2\, 3\rangle^2 \epsilon^{\mu}_{-}(n) \epsilon^{\nu}_{-}(n)\right)  \right) + \mathcal{O}(s_{23}^0)\,,
 \end{split}
\eeq
 or a quark-antiquark pair 
 \beq
 \begin{split}
 g^2 (2 n_f T_F) \frac{2}{(2p_2\cdot p_3)^2} &\left( \left(1-\frac{2 E_2 E_3}{(E_2+E_3)^2}\right) \langle2\, 3\rangle [3\, 2] (\epsilon^{\mu}_{+}(n)\epsilon^{\nu}_{-}(n)+\epsilon^{\nu}_{+}(n)\epsilon^{\mu}_{-}(n)) \right.\\
 &\left. - \frac{2E_2 E_3}{(E_2+E_3)^2} \left( [3\,2]^2 \epsilon^{\mu}_{+}(n)\epsilon^{\nu}_{+}(n)+\langle 2\, 3\rangle^2 \epsilon^{\mu}_{-}(n) \epsilon^{\nu}_{-}(n)\right) 
  \right)+\mathcal{O}(s_{23}^0)\,,
 \end{split}
 \eeq
 where the spinor product $(\langle2\, 3\rangle)^2$ and $(\left[3\, 2\right])^2$ in the second line introduce the azimuthal angle of the splitting plane relative to  some chosen frame in the transverse plane
 \beq
\langle 2 \, 3\rangle=\sqrt{s_{23}} e^{i\phi_S},\quad \left[3\, 2 \right]=\sqrt{s_{23}}e^{-i\phi_S}\,.
 \eeq
Therefore, unlike the case of a quark source, there is an azimuthal dependence in the twist-2 contribution when the source and sink are gluons, which is directly connected to appearance of the transverse spin-2 operators $\mathbb{O}_{\tilde{g},\pm}^{[J]}$ in the OPE
 \beq\label{eq: EE_OPE_gluon_state}
\begin{split}
&\langle 0 | A^{\nu}_b(x) \mathcal{E}(\vec{n}_2)\mathcal{E}(\vec{n}_3) A^\mu_a(0) | 0\rangle\\
=&-\frac{1}{2\pi} \frac{2}{\theta_S^2} \left\{ 
[(\gamma_{gg}(2)-\gamma_{gg}(3))+ 2n_f(\gamma_{qg}(2)-\gamma_{qg}(3)) ] \langle 0| A^\nu_b(x) \mathbb{O}_g^{[3]} A^\mu_a(0)| 0\rangle
\right.\\
&\qquad\qquad +\frac{1}{2} e^{2i\phi_S}  [(\gamma_{g\tilde{g}}(2)-\gamma_{g\tilde{g}}(3))+2n_f(\gamma_{q\tilde{g}}(2)-\gamma_{q\tilde{g}}(3))]
\langle 0| A^\nu_b(x)\mathbb{O}_{\tilde{g},-}^{[3]} A^\mu_a (0) | 0 \rangle \\
& \qquad\qquad \left. 
+\frac{1}{2}e^{-2i\phi_S} [(\gamma_{g\tilde{g}}(2)-\gamma_{g\tilde{g}}(3))+2n_f(\gamma_{q\tilde{g}}(2)-\gamma_{q\tilde{g}}(3))]
 \langle 0| A^\nu_b(x)\mathbb{O}_{\tilde{g},+}^{[3]} A^\mu_a (0) | 0 \rangle 
\right\} +\mathcal{O}(\theta_S^0)\,.
\end{split}
\eeq
Combining (\ref{eq: EE_OPE_quark_state}) and (\ref{eq: EE_OPE_gluon_state}), we can obtain the state-independent $\mathcal{E}\mathcal{E}$ OPE
\beq \label{eq: EE_OPE}
\begin{split}
&\mathcal{E}(\vec{n}_2)\mathcal{E}(\vec{n}_3)=\\
&-\frac{1}{2\pi} \frac{2}{\theta_S^2} \left\{ 
\left[ (\gamma_{qq}(2)-\gamma_{qq}(3))+(\gamma_{gq}(2)-\gamma_{gq}(3))\right]\mathbb{O}_q^{[3]}
+\left[ (\gamma_{gg}(2)-\gamma_{gg}(3))+2 n_f(\gamma_{qg}(2)-\gamma_{qg}(3))\right]\mathbb{O}_g^{[3]}\right.\\
&\qquad\qquad \left. +\frac{1}{2} \left[(\gamma_{g\tilde{g}}(2)-\gamma_{g\tilde{g}}(3))+2 n_f (\gamma_{q\tilde{g}}(2)-\gamma_{q\tilde{g}}(3)) \right]
\left( e^{2i\phi_S} \mathbb{O}_{\tilde{g},-}^{[3]} + e^{-2i\phi_S} \mathbb{O}_{\tilde{g},+}^{[3]}\right)
\right\}+\mathcal{O}(\theta_S^0)\,.
\end{split}
\eeq
As for the case of quark operators, the collinear spin $J=3$ of the light-ray operators appearing in the OPE is fixed by conformal symmetry \cite{Hofman:2008ar}. At higher orders in QCD we expect that we will encounter light-ray operators with non-integer spin in the matching \cite{Dixon:2019uzg}. It will be interesting to understand explicitly how these arise.

\subsubsection{$\mathcal{E}(\vec{n}_1) \vec{\mathbb{O}}^{[J]}(\vec{n}_2)$ OPE}\label{sec:OE_OPE}

For consecutive OPEs, we also need to derive the $\mathcal{E}(\vec{n}_1)\vec{\mathbb{O}}^{[J]}(\vec{n_2})$ OPE. Since we are focusing on the twist-2 branch, we expect the result of this OPE to be a linear combination of the $\vec{\mathbb{O}}^{[J+1]}$ operators, due to the conformal symmetry of QCD at LO.

Schematically, the relevant integrals have the form
\begin{align}
  \label{eq:3}
  \int_0^1 dz \, z^{J-1} (1-z) P_{a \leftarrow b}(z) = \int_0^1 dz\, (z^{J-1} - z^J) P_{a \leftarrow b}(z) \,.
\end{align}
This implies that the OPE result can be better organized if we introduce the OPE coefficient matrix $\widehat{C}_{\phi}(J)$
\begin{align}\label{eq:structure_constants}
\widehat C_\phi(J)&=
\begin{pmatrix}
\gamma_{qq}(J)&&2n_f \gamma_{qg}(J)&&2n_f \gamma_{q\tilde g}(J) e^{-2i\phi}/2 &&2n_f \gamma_{q\tilde g}(J) e^{2i\phi}/2 \\
\gamma_{gq}(J)&& \gamma_{gg}(J)&&\gamma_{g\tilde g}(J) e^{-2i\phi}/2 &&\gamma_{g\tilde g}(J) e^{2i\phi}/2\\
\gamma_{\tilde gq}(J) e^{2i\phi} &&\gamma_{\tilde gg}(J) e^{2i\phi} && \gamma_{\tilde g \tilde g}(J)&&\gamma_{\tilde g \tilde g,\pm}(J) e^{4i\phi}\\
\gamma_{\tilde gq}(J) e^{-2i\phi} &&\gamma_{\tilde gg}(J) e^{-2i\phi} &&\gamma_{\tilde g \tilde g,\pm}(J)e^{-4i\phi}&& \gamma_{\tilde g \tilde g}(J)\\
\end{pmatrix}\,.
\end{align}
where the upper left block is the standard anomalous dimension for unpolarized twist-2 spin-$J$ operators and the phase $e^{i n\phi}$ is included due to the mismatch of the symmetry between the transverse spin-0 operators $\mathbb{O}_q^{[J]},\, \mathbb{O}_g^{[J]}$ and transverse spin-2 operators $\mathbb{O}_{\tilde{g},\pm}^{[J]}$. The convention of the perturbative expansion of $\gamma_{ab}(J)$ in this paper is $\displaystyle \frac{\alpha_s}{4\pi} \gamma_{ab}^{(0)}(J)+\mathcal{O}\left(\alpha_s^2\right)$ and values of $\gamma_{ab}^{(0)}(J)$ are
\beq
\begin{split} \label{eq: gamma_values}
&\gamma_{qq}^{(0)}(J)=C_F\left( 4\left(\psi^{(0)}(J+1)+\gamma_E\right)-\frac{2}{J(J+1)}-3\right),\quad 
\gamma_{qg}^{(0)}(J)=-T_F \frac{2(J^2+J+2)}{J(J+1)(J+2)},\\
&\gamma_{gq}^{(0)}(J)=-C_F \frac{2(J^2+J+2)}{(J-1)J(J+1)}, \quad
\gamma_{gg}^{(0)}(J)= 4 C_A \left( \psi^{(0)}(J+1)+\gamma_E -\frac{1}{(J-1)J}-\frac{1}{(J+1)(J+2)} \right)-\beta_0,\\
&\gamma^{(0)}_{\tilde g \tilde g} (J)=4C_A (\psi ^{(0)}(J+1)+\gamma_E) -\beta_0\,, \quad
\gamma^{(0)}_{\tilde gq}(J)=C_F\frac{2}{(J-1)J}\,,\qquad  \gamma^{(0)}_{\tilde gg}(J)=C_A\frac{2}{(J-1)J}\,, \\
&\gamma^{(0)}_{q\tilde g}(J)= - T_F\frac{8}{(J+1)(J+2)}\,, \qquad \gamma^{(0)}_{\tilde g \tilde g,\pm} (J)=0\,,  \quad
\gamma^{(0)}_{g\tilde g}(J)= C_A \left(\frac{8}{(J+1)(J+2)}+3\right) -\beta_0\,,
\end{split}
\eeq
where $\psi^{(0)}(z) = \Gamma'(z)/\Gamma(z)$ is the digamma function, and $\beta_0 = 11/3 C_A - 4/3 n_f T_F$ is the one-loop beta function in QCD. 

We leave out the details of computation and just state the result for the $\mathcal{E}(\vec{n}_1)\vec{\mathbb{O}}^{[J]}(\vec{n})$ OPE here
\beq
\mathcal{E}(\vec{n}_1) \vec{\mathbb{O}}^{[J]}(\vec{n})=-\frac{1}{2\pi}\frac{2}{\theta_L^2} \left[\widehat{C}_{\phi_L}(J)-\widehat{C}_{\phi_L}(J+1)\right] \vec{\mathbb{O}}^{[J+1]}(\vec{n})+\mathcal{O}(\theta_L^0)\,.
\eeq
In terms of this notation we can also re-organize the $\mathcal{E}(\vec{n}_2)\mathcal{E}(\vec{n}_3)$ given in (\ref{eq: EE_OPE}) since we can infer from mode expansions the relation $\mathcal{E}\sim\mathbb{O}_q+\mathbb{O}_g$. Therefore, we introduce the projection vector 
$\vec{\mathcal{J}}=(1,1,0,0)$ and we have 
\beq
\mathcal{E}(\vec{n}_2)\mathcal{E}(\vec{n}_3)=-\frac{1}{2\pi}\frac{2}{\theta_S^2} \vec{\mathcal{J}}\left[\widehat{C}_{\phi_S}(2)-\widehat{C}_{\phi_S}(3)\right]\vec{\mathbb{O}}^{[3]}(\vec{n})+\mathcal{O}(\theta_S^0)\,.
\eeq

As a result, the leading power of squeezed limit can be obtained from the consecutive OPE
\beq
\mathcal{E}(\vec{n}_1)\mathcal{E}(\vec{n}_2)\mathcal{E}(\vec{n}_3)=\frac{1}{(2\pi)^2} \frac{2}{\theta_S^2}\frac{2}{\theta_L^2}\vec{\mathcal{J}} \left[\widehat{C}_{\phi_S}(2)-\widehat{C}_{\phi_S}(3)\right]\left[\widehat{C}_{\phi_L}(3)-\widehat{C}_{\phi_L}(4)\right] \vec{\mathbb{O}}^{[4]}(\vec{n})+\mathcal{O}(\theta_S^0,\theta_L^0)\,.
\eeq
Here  $\langle \vec{\mathbb{O}}^{[4]}\rangle=(1,0,0,0)^{\mathrm{T}}$ for an unpolarized quark jet and $\langle \vec{\mathbb{O}}^{[4]}\rangle=(0,1,0,0)^{\mathrm{T}}$ for an unpolarized gluon jet. Using the explicit relations for the matching coefficients presented in this section, one can check that one reproduces the leading twist terms in the three-point correlator for both quark and gluons jets (given in \Eqs{eq:quark_result_expand}{eq:gluon_result_expand}). This provides a highly non-trivial check of the light-ray OPE, and of our calculations.

\subsubsection{$ \vec{\mathbb{O}}^{[J_1]}(\vec{n}_1) \vec{\mathbb{O}}^{[J_1]}(\vec{n}_2)$ OPE}\label{sec:OO_OPE}

While for the analysis of the squeezed limit of the three-point correlator only the  $\mathcal{E}(\hat n_1) \mathcal{E}(\hat n_2)$ and $\mathcal{E}(\hat n_1) \mathbb{O}^{[J]}(\hat n_2) $ OPEs are required,  squeezed limits of higher point correlators can be studied in an identical manner. The only additional OPE that is required for the study of squeezed limits of generic correlators is the  $\mathbb{O}^{[J_1]}(\hat n_1) \mathbb{O}^{[J_2]}(\hat n_2) $ OPE. An interesting case where this appears is the 4-point energy correlator $\langle\mathcal{E}(\vec{n}_1) \mathcal{E}(\vec{n}_2)\mathcal{E}(\vec{n}_3)\mathcal{E}(\vec{n}_4) \rangle$ in a pairwise squeezed limit $\theta_{12},\theta_{34}\ll \theta_{13}\ll 1$. For completeness we therefore present the additional $\mathbb{O}^{[J_1]}(\hat n_1) \mathbb{O}^{[J_2]}(\hat n_2) $ in this section. This completes the set of OPEs required to study squeezed limits of arbitrary correlators in QCD jets at leading twist.

Schematically, the relevant integrals have the form
\begin{align}
  \label{eq:4}
  \int_0^1 dz\, z^{J_1 - 1} (1-z)^{J_2 - 1} P_{a \leftarrow b}(z) \,.
\end{align}
Writing the splitting function with inverse Mellin transformation,
\begin{align}
  \label{eq:5}
  P_{a \leftarrow b}(z) = - \frac{1}{2} \int_{c - i \infty}^{c + i \infty} \frac{d J}{2 \pi i} z^{-J} \gamma_{ab}(J) \,,
\end{align}
the $z$ integral in \eqref{eq:4} can be done in closed form. 
We then find that the OPEs of twist-2 spin-$J$ operators in QCD are given by
\begin{align}
\mathbb{O}_{q}^{[J_1]}(\vec{n}_1) \mathbb{O}_{g}^{[J_2]}(\vec{n}_2)
&=-\frac{1}{2\pi} \frac{2}{\theta^2}\left( \int \frac{dJ}{2\pi i} B(J_1-J,J_2) \gamma_{qq}(J)\right) \mathbb{O}_{q}^{[J_1+J_2-1]}+\text{higher twist}\,,\nonumber\\
\mathbb{O}_{q}^{[J_1]}(\vec{n}_1) \mathbb{O}_{q}^{[J_2]}(\vec{n}_2)
&= -\frac{1}{2\pi}\frac{2}{\theta^2} \left[2n_f \left( \int\frac{dJ}{2\pi i} B(J_1-J,J_2) \gamma_{qg}(J) \right)\mathbb{O}_{g}^{[J_1+J_2-1]}\right. \nonumber\\
&\hspace{-3cm} \; \left. + 2n_f \left( \int \frac{dJ}{2\pi i} B(J_1-J,J_2) \gamma_{q\tilde{g}}(J) \right) 
\left(\frac{e^{2i\phi}}{2}\mathbb{O}_{\tilde{g},-}^{[J_1+J_2-1]}+\frac{e^{-2i\phi}}{2}\mathbb{O}_{\tilde{g},+}^{[J_1+J_2-1]}\right)   \right]
+\text{higher twist}\,,\nonumber\\
\mathbb{O}_{g}^{[J_1]}(\vec{n}_1) \mathbb{O}_{g}^{[J_2]}(\vec{n}_2)
&=-\frac{1}{2\pi}\frac{2}{\theta^2}\left[ \left(\int \frac{dJ}{2\pi i} B(J_1-J,J_2)\gamma_{gg}(J) \right) \mathbb{O}_{g}^{[J_1+J_2-1]}\right. \nonumber\\
&\hspace{-3cm} \; \left. + 2n_f \left( \int \frac{dJ}{2\pi i} B(J_1-J,J_2) \gamma_{g\tilde{g}}(J) \right) 
\left(\frac{e^{2i\phi}}{2}\mathbb{O}_{\tilde{g},-}^{[J_1+J_2-1]}+\frac{e^{-2i\phi}}{2}\mathbb{O}_{\tilde{g},+}^{[J_1+J_2-1]}\right)   \right]
+\text{higher twist}\,,\nonumber\\
\mathbb{O}_{\tilde{g},\pm}^{[J_1]}(\vec{n}_1)\mathbb{O}_{q}^{[J_2]}(\vec{n}_2)
&= -\frac{1}{2\pi} \frac{2}{\theta^2} \left( \int \frac{dJ}{2\pi i} B(J_1-J,J_2)\gamma_{\tilde{g}q}\right) e^{\pm 2i\phi} \mathbb{O}_{q}^{[J_1+J_2-1]} +\text{higher twist}\,, \nonumber\\
\mathbb{O}_{\tilde{g},\pm}^{[J_1]}(\vec{n}_1)\mathbb{O}_{g}^{[J_2]}(\vec{n}_2)
&=-\frac{1}{2\pi}\frac{2}{\theta^2}\left[ \left(\int\frac{dJ}{2\pi i} B(J_1-J,J_2) \gamma_{\tilde{g}\tilde{g}}(J)\right)\mathbb{O}_{\tilde{g},\pm}^{[J_1+J_2-1]} \right. \nonumber\\
& \hspace{-3cm} \left. +    \left(\int\frac{dJ}{2\pi i} B(J_1-J,J_2) \gamma_{\tilde{g}g}(J)\right)e^{\pm 2i\phi}\mathbb{O}_{g}^{[J_1+J_2-1]}    \right]+\text{higher twist}\,.
\end{align}
Here $B(x,y) = \Gamma(x) \Gamma(y)/\Gamma(x+y)$ is the standard Beta function, and the contour is taken along the imaginary axis, lying to the R.H.S. of all the poles of $\gamma_{ab}(J)$ and to the L.H.S. of all the poles of $B(J_1 -J,J_2)$, as is standard in Barnes integrals. 
The $\gamma_{ab}(J)$ introduced in (\ref{eq: gamma_values}) is not enough to describe $\mathbb{O}_{\tilde{g},\lambda_1}^{[J_1]}\mathbb{O}_{\tilde{g},\lambda_2}^{[J_2]}$ OPE because both operators need to be polarized. Therefore, we further define
\bea
\gamma_{+-\leftarrow g}(J)&=&\frac{\alpha_s}{4\pi}\left(-2C_A \frac{1}{(J+1)(J+2)}\right)+\mathcal{O}(\alpha_s^2)\,,\nonumber\\
\gamma_{+-\leftarrow +}(J)&=&\frac{\alpha_s}{4\pi} \left(-2C_A(\psi^{(0)}(J+3)+\gamma_E)\right)+\mathcal{O}(\alpha_s^2)\,,\nonumber\\
\gamma_{+-\leftarrow -}(J)&=&\frac{\alpha_s}{4\pi}\left( C_A \frac{12}{(J-1) J (J+1) (J+2)}\right)+\mathcal{O}(\alpha_s^2)\,,\nonumber\\
\gamma_{++\leftarrow +}(J)&=&\frac{\alpha_s}{4\pi}\left(-2 C_A(\psi^{(0)}(J-1)+\gamma_E)  \right)+\mathcal{O}(\alpha_s^2)\,,\nonumber\\
\gamma_{++\leftarrow g}(J)&=&\gamma_{++\leftarrow -}=\mathcal{O}(\alpha_s^2)\,,\nonumber
\eea
and write $\mathbb{O}_{\tilde{g},\lambda_1}^{[J_1]}\mathbb{O}_{\tilde{g},\lambda_2}^{[J_2]}$ OPE in a similar form
\bea
& &\mathbb{O}_{\tilde{g},+}^{[J_1]}(\vec{n}_1)\mathbb{O}_{\tilde{g},-}^{[J_2]}(\vec{n}_2)
= -\frac{1}{2\pi} \frac{2}{\theta^2}\left[ \left(\int \frac{dJ}{2\pi i} B(J_1-J,J_2) \gamma_{+-\leftarrow +}(J)\right) e^{-2i\phi} \mathbb{O}_{\tilde{g},+}^{[J_1+J_2-1]}   \right.\nonumber\\
& & \left. +\left(\int \frac{dJ}{2\pi i} B(J_1-J,J_2)\gamma_{+-\leftarrow -}(J) \right)e^{2i\phi}\mathbb{O}_{\tilde{g},-}^{[J_1+J_2-J]} +\left(\int \frac{dJ}{2\pi i}B(J_1-J,J_2)\gamma_{+-\leftarrow g}(J)\right)\mathbb{O}_{g}^{[J_1+J_2-J]}\right] \nonumber\\
& & +\text{higher twist}\,, \nonumber\\
& &\mathbb{O}_{\tilde{g},+}^{[J_1]}(\vec{n}_1)\mathbb{O}_{\tilde{g},+}^{[J_2]}(\vec{n}_2)
= -\frac{1}{2\pi} \frac{2}{\theta^2}\left[ \left(\int \frac{dJ}{2\pi i} B(J_1-J,J_2) \gamma_{+-\leftarrow +}(J)\right) e^{2i\phi} \mathbb{O}_{\tilde{g},+}^{[J_1+J_2-1]}   \right.\nonumber\\
& & \left. +\left(\int \frac{dJ}{2\pi i} B(J_1-J,J_2)\gamma_{++\leftarrow +}(J) \right)e^{6i\phi}\mathbb{O}_{\tilde{g},+}^{[J_1+J_2-J]} +\left(\int \frac{dJ}{2\pi i}B(J_1-J,J_2)\gamma_{++\leftarrow g}(J)\right)e^{4i\phi}\mathbb{O}_{g}^{[J_1+J_2-J]}\right]\nonumber \\
& & +\text{higher twist}\,. \nonumber\\
\eea
It will be interesting to test these results when perturbative data for the four-point correlator is available.

\subsection{Higher-Point OPEs at Leading Twist}\label{sec:comments}

While the goal of the light-ray OPE is a convergent expansion, for many phenomenological purposes in QCD, even the leading power information is useful. Indeed, for nearly all applications of resummation at the LHC, only leading power information is used. Since we have seen how the light-ray OPE provides a convenient organization for the OPE of two light-ray operators, it is interesting to consider the simultaneous OPE of multiple light-ray operators at leading twist.

We consider a multi-point OPE of some set of light-ray operators, and make the assumption that there exists an OPE onto a single light-ray operator. At leading twist, the light-ray operators in a perturbative gauge theory are known and are classified by their field content, and their quantum numbers collinear spin $J$ and transverse spin $j$. In particular, in QCD, they are given by the operators $\mathbb{O}_i^{[J]}$. It therefore suffices to determine the matching coefficient, as was done in the previous sections for the $\cE(\vec n_1) \cE(\vec n_2)$ OPE. However, the key point is that this is precisely equivalent to the calculation of the multipoint correlator in the collinear limit, as was performed explicitly for the three-point correlator in  \cite{Chen:2019bpb}. In particular, extending the calculation of \Sec{sec:EE_OPE} to the calculation of an $k$ point correlator in the collinear limit, one can again perform the change of variables as in \Eq{eq:change_var}, $E_i \to z_i E$. The integral over $E$ reproduces the behavior of the light-ray operator with spin $k+1$, while the integral over the energy fractions reproduces the shape dependence of the multipoint correlator. 

Therefore, at leading order, and leading twist, we obtain the following OPE for a $k$ point correlator onto leading twist light-ray operators
\begin{align} \label{eq:general_OPE}
\mathcal{E}(\vec{n}_1)\mathcal{E}(\vec{n}_2)\cdots  \mathcal{E}(\vec{n}_k)
&= \frac{1}{\theta_L^2} \left\{ f^{[k]}_q(u_i,v_i) \mathbb{O}_q^{[k+1]}(n_1)+ f^{[k]}_g(u_i,v_i) \mathbb{O}_g^{[k+1]}(n_1) \right.  \\
&\left. + f^{[k]}_{g+}(u_i,v_i)  e^{2i\phi_S} \mathbb{O}_{\tilde{g},-}^{[k+1]}(n_1) + f^{[k]}_{g-}(u_i,v_i)  e^{-2i\phi_S} \mathbb{O}_{\tilde{g},+}^{[k+1]}(n_1)
\right\}
+\mathcal{O}(\theta_L^0) \nn\,.
\end{align}
Here $f^{[k]}_j(u_i,v_i)$ are functions of the conformal cross-ratios $u_i,v_i$, and $\theta_L$ describes the overall size, and can be taken to be the longest side. This should be compared with the form of the three-point function given in \Eq{eq:three_point_general}. The functions $f^{[3]}_g(u,v)$ and $f^{[3]}_q(u,v)$ were explicitly calculated in \cite{Chen:2019bpb}, and are given in \Eq{eq:three_point_N4} for $\cN=4$ SYM. The appearance of light-ray operators with fixed spin $k=1$ is due to conformal symmetry \cite{Hofman:2008ar,Kologlu:2019mfz}. In QCD this is only expected to leading order, but this is sufficient to perform leading logarithmic resummation. This result also agrees with the original statement made by Hofman and Maldacena \cite{Hofman:2008ar}  that the scaling is independent of the shape, and our argument is essentially the perturbative version of their statement.

The result of \Eq{eq:general_OPE} is particularly interesting for phenomenological applications where one must perform small angle resummation. Under the assumption of the light-ray OPE, it reduces the scaling behavior of an arbitrarily shaped energy correlator to the scaling behavior of a fixed set of light-ray operators. The anomalous dimensions of these light-ray operators in QCD will be calculated as a function of $J$ in \Sec{sec:resum}, which therefore allows the prediction of the scaling properties of arbitrarily shaped correlators at the LHC. It will be interesting to test this scaling both with higher order calculations and in LHC data.

\section{Resummation in the Squeezed Limit}\label{sec:resum}

For phenomenological applications, for example the measurement of the three-point correlator at the LHC, one must perform an all-order resummation of large logarithms that appear in the squeezed limit. These logarithms arise due to the presence of the three different energy scales that appear in the squeezed limit, namely the jet energy scale $Q$, the energy scale of the first splitting $\theta_L Q$ and the energy scale of the second splitting $\theta_S Q$. The use of light-ray operators offers a clear way to perform the resummation of these logarithms, including a proper incorporation of all transverse spin effects, through the renormalization of the light-ray operators appearing in the OPE.

Under RG evolution, $\mathcal{O}_q^{[J]}$ and $\mathcal{O}_g^{[J]}$ cannot mix with $\mathcal{O}_{\tilde{g},\pm}^{[J]}$ since they have different transverse spins. The anomalous dimension of the unpolarized twist-2 operators $(\mathcal{O}_q^{[J]}, \mathcal{O}_g^{[J]})$ are well known, and were calculated  in the original work of Gross and Wilczek \cite{Gross:1974cs,Gross:1973id,Gross:1973ju},  so here we only present the calculation of the transverse spin-2 anomalous dimensions. We define $O_{J}^{ij}=-2^J \mathcal{O}_{\tilde{g}}^{[J],ij}$ for simplicity since the prefactor is irrelevant for the anomalous dimension calculation.

To perform the calculation we use matrix element with gluonic external states 
\beq
\langle k,+| O_{J, \lambda} |k,-\rangle \quad \text{or} \quad \langle k,-| O_{J,\lambda} |k,+\rangle\,,
\eeq
where
\begin{align}
O_{J,\lambda}= \epsilon_{\lambda,i}\epsilon_{\lambda,j} O_{J}^{ij}= \epsilon_{\lambda,i}\epsilon_{\lambda,j} F^{+i}(iD^+)^{J-2}F^{+j}\,.
\end{align}
As long as we constrain ourselves to the twist-2 operators, we can take the on-shell limit $k^2\to 0$ because divergences that contains $k^2$ contribute to the mixing with higher twist operators. We denote the external polarization vector by $\tilde{\epsilon}_{\lambda^\prime}$ which satisfies $\tilde{\epsilon}_{\lambda^\prime}^2=k\cdot\tilde{\epsilon}_{\lambda^\prime}=0$.

\begin{figure}
\centering
\subfloat[]{\includegraphics[width=3cm]{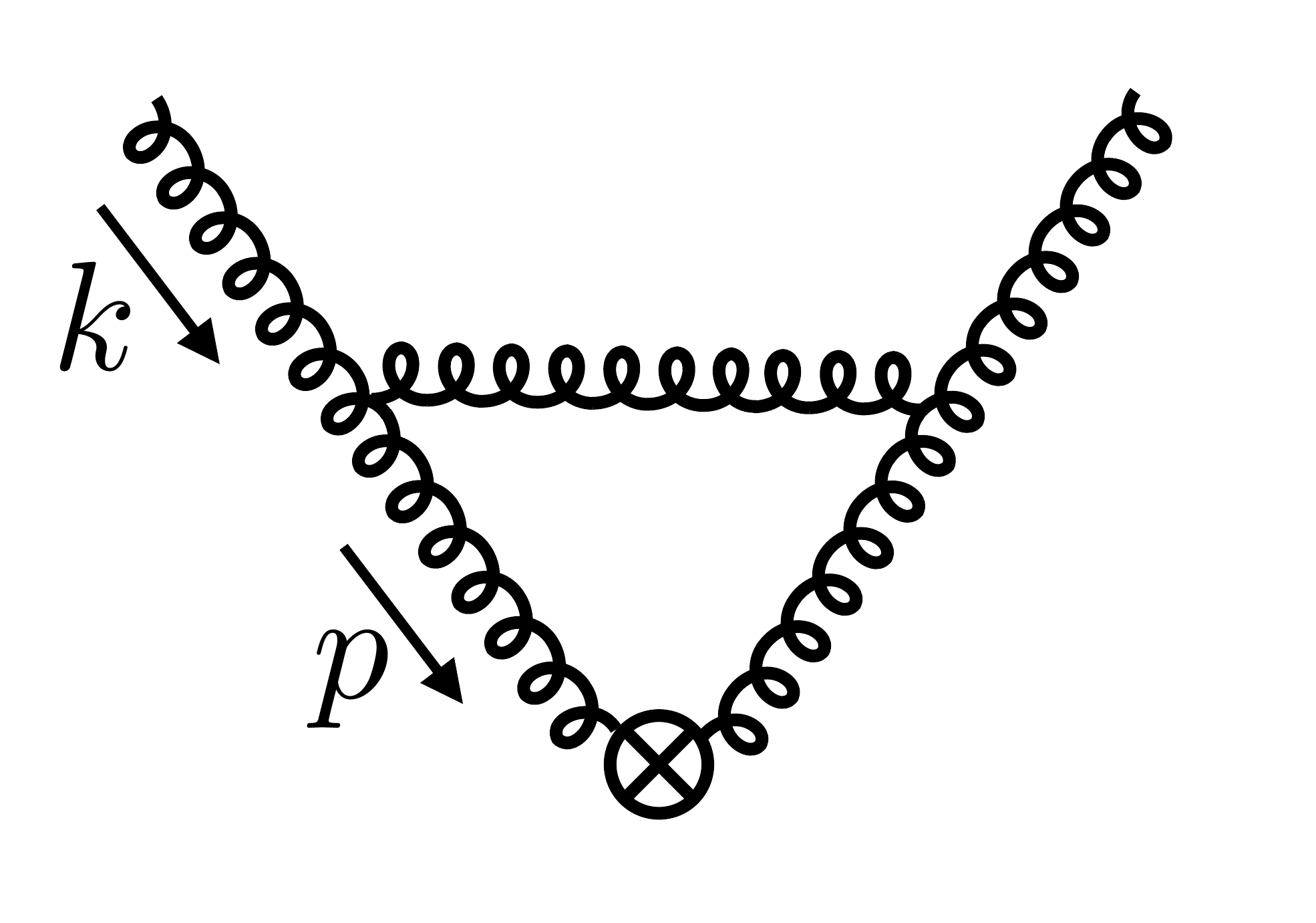} \label{fig: triangle}}
\subfloat[]{\includegraphics[width=6cm]{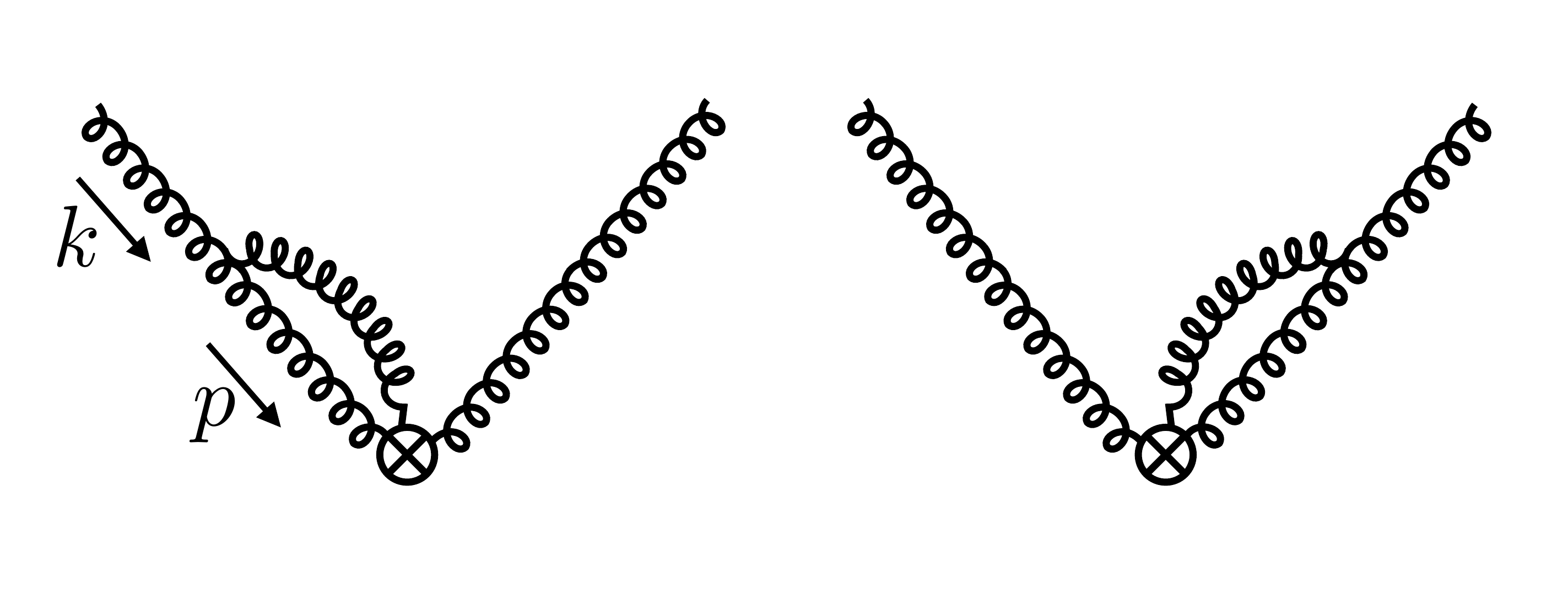} \label{fig: bubble}}\\
\caption{Feynman diagrams contributing to the one-loop anomalous dimensions of transverse spin-2 operators, as described in the text.}
\label{fig: Feynman_Graphs}
\end{figure}

Two kinds of graphs contribute at 1 loop order (Fig.\ref{fig: Feynman_Graphs} \subref{fig: triangle}\subref{fig: bubble}). The two- and three-gluon Feynman rules for the vertex $O_{J}^{ij}$ are respectively
\begin{align}\label{eq: Feynman_Rule1}
\hspace{-1cm}\fd{2.3cm}{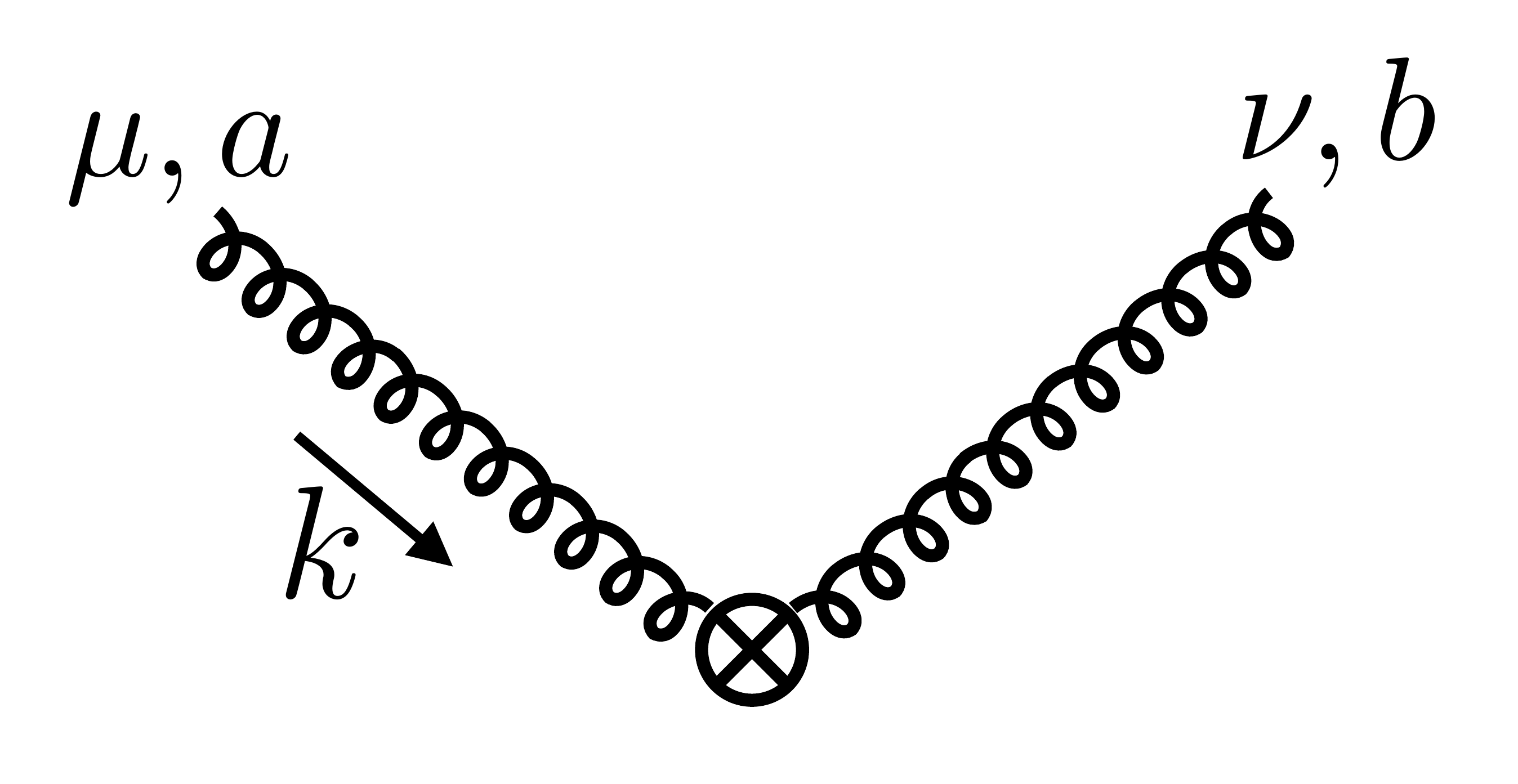}&=\delta_{ab}( ik^{+} g^{\mu i}-ik^{i}g^{\mu +})(k^+)^{J-2}(-ik^+ g^{\nu j}+ik^{j}g^{\nu +})+\left(k\to -k, \mu\leftrightarrow \nu\right)\,,
\end{align}
and
\begin{align}\label{eq: Feynman_Rule2}
\fd{2.4cm}{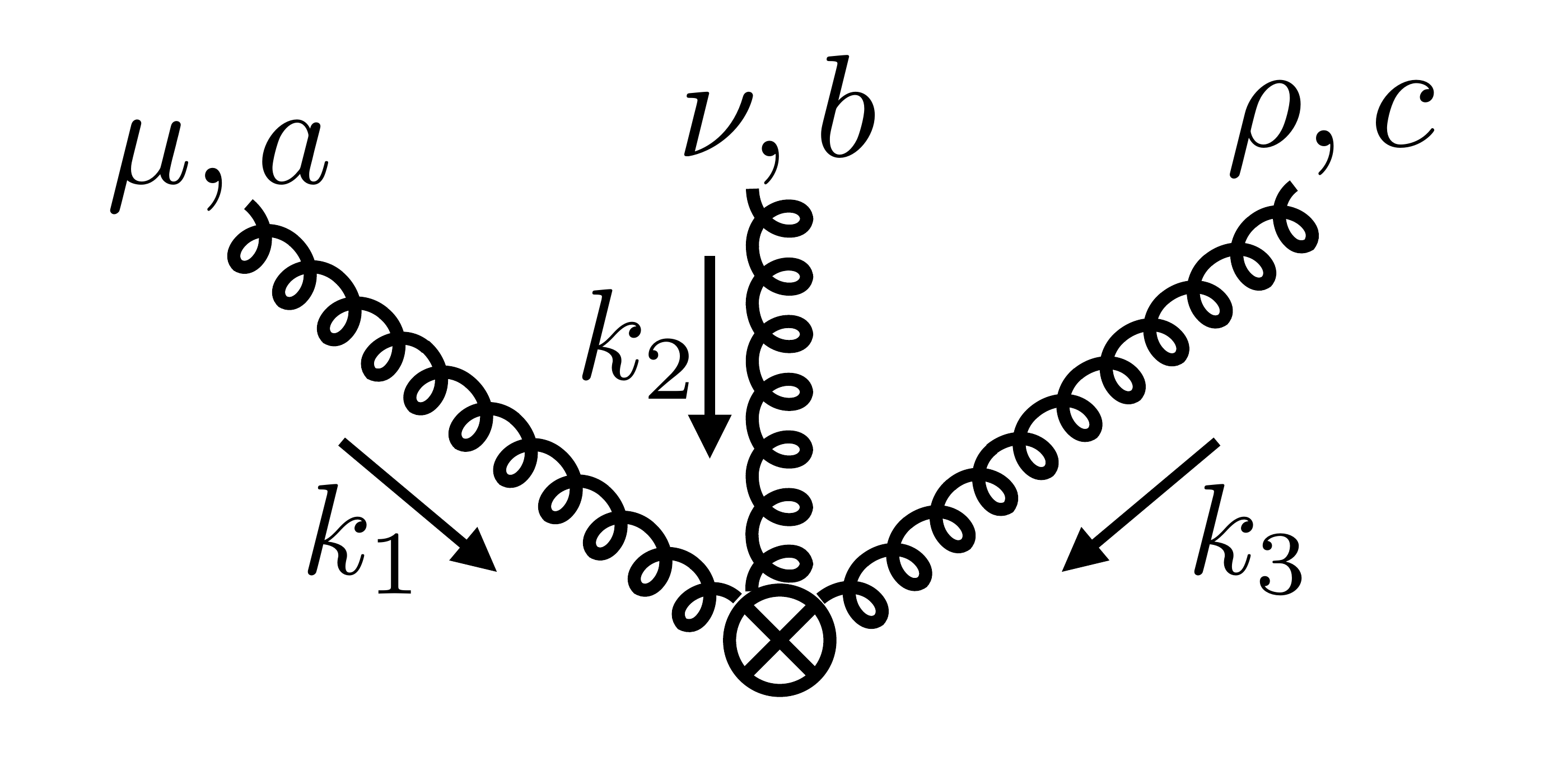}&=g f_{abc} g^{\nu +}g^{\rho i}(k_1^+)^{J-2}(-ik_1^{+}g^{\mu j}+ik_1^j g^{\mu +})\nn \\
&+g f_{abc}(-ik_1^{+}g^{\mu i}+ik_1^i g^{\mu +})(k_2^+ + k_3^+)^{J-2}g^{\nu +}g^{\rho j} \nn \\
&+i g f_{abc}\sum_{\alpha=2}^{J-1} (-i k_1^+ g^{\mu i}+i k_1^{i} g^{\mu +}) (k_2^+ +k_3^+)^{\alpha-2} g^{\nu +} (k_3^{+})^{J-1-\alpha}(-ik_3^+ g^{\rho j}+i k_3^{j}g^{\rho +})\nn\\
&+(\mbox{permutations on } \{\mu a k_1\}, \{\nu b k_2\},\{\rho c k_3 \})\,.
\end{align}
From (\ref{eq: Feynman_Rule1}), we can immediately write down the tree level matrix element by contracting it with $\epsilon_{\lambda,i}\epsilon_{\lambda,j}\tilde{\epsilon}_{\lambda^\prime,\mu}\tilde{\epsilon}_{\lambda^\prime,\nu}$,
\beq
2\delta_{ab} \left( 
(\epsilon_\lambda \cdot \tilde{\epsilon}_{\lambda^\prime})^2 (\bar{n}\cdot k)^{J} -2 (\epsilon_\lambda\cdot\tilde{\epsilon}_{\lambda^\prime})(\bar{n}\cdot k)^{J-1} + (\bar{n} \cdot \tilde{\epsilon}_{\lambda^\prime})^2 (k \cdot \epsilon_\lambda)^2 (\bar{n}\cdot k)^{J-2}
\right)\,.
\eeq
We emphasize again that this result is obtained from the analytic continuation of the even $J$ branch. Since the  transverse spin-2 operators cannot mix with the transverse spin-0 twist-2 operators and since we can neglect mixing with higher twist operators, we expect the same structure will appear in the numerator of the divergent term of the 1-loop calculation. We can therefore simplify our calculation by focusing on the term that contains $(\bar{n}\cdot k)^J$ when extracting ingredients for the anomalous dimension calculation.

For the diagram in Fig.\ref{fig: Feynman_Graphs}\subref{fig: triangle}, the relevant loop integral is
\beq
I_1^{\mu\nu;ij}=\int \frac{d^d p}{(2\pi)^d} \frac{-i}{(k-p)^2}\left( \frac{-i}{p^2}\right)^2\times N_1^{\mu\nu;ij}=\int_0^1 dx \int \frac{d^d q}{(2\pi)^d} \frac{2i(1-x) N_1^{\mu\nu;ij}|_{p\to q+x k}}{(q^2+(1-x)x k^2)^3}\,,
\eeq
where $\mu,\nu$ are Lorentz indices of external gluons while $i,j$ refers to the indices of vertex $O_{J}^{ij}$ and $N_1^{\mu\nu;ij}$ is a complicated numerator depending on $k,p$. The logarithmic divergence gives rise to the leading twist anomalous dimension, so we only need quadratic terms of $q$ in $N_1^{\mu\nu,ij}|_{p\to q+ xk}$.  After some manipulation, the coefficient of $2\delta_{ab} (\epsilon_\lambda\cdot\tilde{\epsilon}_{\lambda^\prime})^2 (\bar{n}\cdot k)^J q^2$ in $N_1$ is found to be
\beq
-\frac{1}{2} g^2 C_A \left(J x^{J-1}+(J+2)x^J \right) \label{eq: triangle_coefficient}\,.
\eeq
Similarly, for the diagrams in Fig.\ref{fig: Feynman_Graphs}\subref{fig: bubble}, we have
\beq
I_2^{\mu\nu;ij}=\int \frac{d^d p}{(2\pi)^d} \frac{-i}{(k-p)^2} \frac{-i}{p^2}\times N_2^{\mu\nu;ij}=-\int_0^1 dx \int \frac{d^d q}{(2\pi)^d} \frac{ N_2^{\mu\nu;ij}|_{p\to q+x k}}{(q^2+(1-x)x k^2)^2}\,,
\eeq
where the $q$ independent terms in $N_2^{\mu\nu;ij}|_{p\to q+ xk}$ contribute to the logarithmic divergence. We find that the coefficient of $2\delta_{ab}(\epsilon_\lambda\cdot\tilde{\epsilon}_{\lambda^\prime})^2 (\bar{n}\cdot k)^J$ is
\beq
-2 i g^2 C_A \left(\sum_{n=1}^{J} \frac{1}{n}+ \sum_{n=1}^{J+1}\frac{1}{n}-1 \right)\,. \label{eq: bubble_coefficient}
\eeq
 Therefore, combining (\ref{eq: triangle_coefficient}), (\ref{eq: bubble_coefficient}) and the integrals
\bea
\left. \int \frac{d^{d}q}{(2\pi)^d} \frac{i q^2}{(q^2+(1-x)x k^2)^3}\right|_{d\to 4-2\epsilon}=-\frac{1}{(4\pi)^2\epsilon}+\mathcal{O}(\epsilon^0)\,,\\
\left. -\int \frac{d^d q}{(2\pi)^d} \frac{1}{(q^2+(1-x)x k^2)^2}\right|_{d\to 4-2\epsilon}=-\frac{i}{(4\pi)^2\epsilon}+\mathcal{O}(\epsilon^0)\,,
\eea
we find that the 1-loop logarithmic divergence is cancelled by the counter-term $\delta_{O_J} O_{J,\lambda}$ with
\beq
\delta_{O_J}= \frac{g^2}{(4\pi)^2\epsilon} 2C_A \left(2\sum_{n=1}^J\frac{1}{n}-1\right)\,.
\eeq
In order to obtain the anomalous dimension, we have to further subtract the field strength renormalization counter-term $\displaystyle \delta_3= \frac{g^2}{(4\pi)^2 \epsilon} \left( \frac{5}{3} C_A - \frac{4}{3}n_f T_F \right)$ since $O_{J,\lambda}$ contains 2 gluon fields at LO. As a result, the RG evolution of twist-2 spin-$J$ transverse spin-2 operator is given by
\beq
\frac{d}{d\ln \mu^2} O_{J,\lambda} = -\gamma_{\tilde{g}\tilde{g}}(J) O_{J,\lambda}, \quad \text{where}\; \gamma_{\tilde{g}\tilde{g}}(J)=\frac{g^2}{(4\pi)^2} \left(4C_A \sum_{n=1}^{J} \frac{1}{n}-\beta_0 \right)\,.
\eeq

We can then promote the RG equation for the local operators to an RG equation for the light-ray operators
\beq
\frac{d}{d\ln \mu^2}\vec{\mathbb{O}}^{[J]}=-\hat{\gamma}(J)\cdot \vec{\mathbb{O}}^{[J]}\,,
\eeq
where 
\begin{align}\label{eq:anom_dim}
\hat \gamma(J)=
\begin{pmatrix}
\gamma_{qq}(J)&&2n_f \gamma_{qg}(J)&&0\\
\gamma_{gq}(J)&& \gamma_{gg}(J)&&0\\
0&&0&& \gamma_{\tilde g \tilde g}(J)\mathbf{1}
\end{pmatrix}\,,
\end{align}
with $\mathbf{1}$ a $2\times 2$ identity matrix. After solving RG equation, we can resum the large logarithms for the leading power expansion of the squeezed limit to obtain a resummed prediction for the three-point correlator
\begin{align}
\mathcal{E}(\hat n_1) \mathcal{E}(\hat n_2) \mathcal{E}(\hat n_3)
&=\frac{1}{(2\pi)^2} \frac{2}{\theta_S^2}\frac{2}{\theta_L^2} \vec {\cal J} \left[ \widehat C_{\phi_S}(2)- \widehat C_{\phi_S}(3)  \right] \left[ \frac{\alpha_s(\theta_L Q)}{\alpha_s(\theta_S Q)}\right]^{\frac{\widehat \gamma^{(0)}(3)}{\beta_0}}  \nn \\
&\times \left[ \widehat C_{\phi_L}(3)-\widehat C_{\phi_L}(4)  \right]  \left[ \frac{\alpha_s(Q)}{\alpha_s(\theta_L Q)}\right]^{\frac{\widehat \gamma ^{(0)}(4)}{\beta_0}} 
\vec {\mathbb{O}}^{[4]} (\hat n_1) 
+\cdots\,,
\end{align}
where the dots denote higher twist terms, as well as higher order terms in the coupling expansion
This final result was used in \cite{Chen:2020adz} to make predictions for the resummed squeezed limit for quark and gluon jets at the LHC, but here we have presented a systematic derivation using the light-ray OPE.

Here we have focused on the leading twist behavior of the squeezed limit of the three-point function, but we believe that our results can be extended in a number of ways. First, we have computed the anomalous dimensions $\gamma_{\tilde g \tilde g}(J)$ for generic values of $J$, which will appear in the squeezed limits of higher point corelators. We therefore believe that these anomalous dimensions have widespread applicability to the study of multipoint energy correlators in jet substructure. Second, we believe that the entire set of subleading twist corrections with highest transverse spin should also resum with this same anomalous dimension (potentially up to $\beta$-function corrections). This would also be interesting to prove directly, as well as to understand how to include corrections from the breaking of conformal symmetry. Finally, it will be interesting to extend these results to higher logarithmic accuracy. Using the light-ray OPE formalism, this requires understanding reciprocity \cite{Basso:2006nk,Chen:2020uvt} for the EEC \cite{Dixon:2019uzg}.

\section{Conclusions}\label{sec:conc}

In this paper we have studied in detail the squeezed (OPE) limit of the three-point energy correlator, focusing in particular on its transverse spin structure, and its description in terms of light-ray operators.  The goal of this paper was both to understand the structure of the OPE limit, as well as to develop the light-ray OPE in QCD at leading twist. 

We computed the $\mathcal{E}(\hat n_1) \mathcal{E}(\hat n_2)$, $\mathcal{E}(\hat n_1) \mathbb{O}^{[J]}(\hat n_2)$ and $\mathbb{O}^{[J_1]}(\hat n_1) \mathbb{O}^{[J_2]}(\hat n_2) $  light-ray OPEs as analytic functions of $J$, and showed that this OPE closes at leading twist in QCD. The light-ray OPE therefore provides a convenient approach for the study of the squeezed limits of light-ray operators, particularly in the presence of transverse spin. For the particular case of $J=3$, we used our results to reproduce the leading power terms in the expansion of the three-point correlator calculated analytically in previous work. We find exact agreement, which provides a highly non-trivial test of the light-ray OPE. 
By computing the anomalous dimensions of the light-ray operators, we performed the resummation of the dominant leading twist terms in the squeezed limit of the three-point corelator. Finally, we showed that this same matching procedure can also be used to perform the OPE of multiple light-ray operators, and we interpreted our previous calculation of the three-point energy correlator in terms of the light-ray OPE.  This is important for phenomenological applications of multi-point correlators for jet substructure at the LHC, and allows their use for precision studies of QCD.

We also made several comments on the higher twist structure of the squeezed limit. In particular, we noted that the highest transverse spin contributions in QCD are described by a single celestial block, and do not receive contributions from higher twist primaries. This enabled us to directly check the form of celestial blocks using our perturbative data. A complete analysis of the celestial blocks, and the higher twist structure of the correlator will be presented in future work  \cite{blocks:forthcoming}. Second, we noted the presence of subleading twist terms proportional to $\log(2|w|)$, where $w$ is the parameter that controls the approach to the OPE limit. From the perspective of perturbative calculations, these arise when the particles going into the detectors that are not squeezed become soft, and are similar to ``endpoint divergences" appearing in QCD factorization. It would be interesting to better understand this relation, and if the light-ray OPE can provide insight into the generic endpoint divergences in gauge theory factorization.

On the more phenomenological side, we believe that the study of spin correlations in jet substructure, for which squeezed limits of energy correlators present particularly clean observables to study, merits more efforts, both in data, and in comparison to parton shower generators. Parton shower generators often neglect spin correlation effects in the shower, which can lead to significant errors for jet substructure observables that are sensitive to multi-point structures \cite{Dasgupta:2020fwr}. There has recently been renewed interest in the development of improved parton showers, including the inclusion of algorithms for treating spin interference effects \cite{Collins:1987cp,Knowles:1988hu,Knowles:1988vs,Knowles:1987cu,Richardson:2001df,Richardson:2018pvo} in Herwig \cite{Richardson:2018pvo,Bellm:2019zci}. Due to the analytic control over the squeezed limits of the energy correlators, they could provide important observables for testing these parton shower generators. Indeed, as this paper was being finalized, \cite{1854526} appeared, in which spin correlations were implemented in the PanScales \cite{Dasgupta:2020fwr,Hamilton:2020rcu}  family of parton showers, and the squeezed limit of the three-point energy correlator, along with the analytic results presented here and in \cite{Chen:2020adz}, was used as a validation of the accuracy of the shower. We hope for more such fruitful interplay between analytic calculations and parton shower development in the future.

We believe that the light-ray OPE shows promise as a widely applicable theoretical technique for the study of jet substructure. The results of this paper, namely the development of the light-ray OPE in QCD at leading twist is a first step in its applications beyond the context of CFTs. We hope to see significant advances in its understanding and applications in the future.

\begin{acknowledgments}

We thank Cyuan-Han Chang, Zhong-Jie Huang, David Simmons-Duffin, Xiaoyuan Zhang, and Alexander Zhiboedov for useful discussions, and to Joshua Sandor for useful discussions and collaboration on related topics. 
HC and HXZ were supported by National Science Foundation of China under contract No.~11975200.  
I.M was supported by the Office of High Energy Physics of the U.S. Department of Energy under Contract No. DE-AC02-76SF00515. HXZ is grateful to the support from the Qiushi Science and Technologies Foundation.

\end{acknowledgments}

\bibliography{EEC_forward}{}
\bibliographystyle{JHEP}

\end{document}